\documentclass[twocolumn,preprintnumbers,prd,superscriptaddress,nofootinbib,floatfix,showpacs,showkeys]{revtex4-1}
\usepackage{upgreek}

\usepackage{amsmath}
\usepackage{amsfonts}
\usepackage{amssymb}
\usepackage{graphicx}
\usepackage[titletoc]{appendix}
\usepackage[normalem]{ulem}
\usepackage{cancel}
\usepackage{color}
\usepackage[rightcaption]{sidecap}
\usepackage{subfigure}
\usepackage{comment}
\usepackage{mathtools}
\usepackage{dcolumn}
\usepackage{array}
\usepackage{ctable}
\usepackage{multirow}
\usepackage{siunitx}
\usepackage{longtable}
\usepackage{tabularx}
\usepackage{booktabs}

\graphicspath{{Graphics/}}
\def\be{\begin{equation}}
\def\ee{\end{equation}}
\def\bea{\begin{eqnarray}}
\def\eea{\end{eqnarray}}
\newcommand{\onetwo}{\frac{1}{2}}

\newcommand{\onefour}{\frac{1}{4}}
\newcommand{\dop}{\dot{\phi}}
\newcommand{\ddop}{\ddot{\phi}}

\newcommand{\PR}{\mathcal{P}_\mathcal{R}}
\newcommand{\PT}{\mathcal{P}_T}

\newcommand{\FF}{F^{\mu\nu}F_{\mu\nu}}
\newcommand{\MpXi}{\left(M_P^2-\xi\phi^2\right)}

\definecolor{vividviolet}{rgb}{0.62, 0.0, 1.0}
\definecolor{amaranth}{rgb}{0.9, 0.17, 0.31}
\definecolor{palatinateblue}{rgb}{0.15, 0.23, 0.89}
\definecolor{brightpink}{rgb}{1.0, 0.0, 0.5}
\definecolor{cornflowerblue}{rgb}{0.39, 0.58, 0.93}
\definecolor{deepcarminepink}{rgb}{0.94, 0.19, 0.22}
\definecolor{radicalred}{rgb}{1.0, 0.21, 0.37}

\usepackage{hyperref}
\usepackage{cleveref}
\hypersetup{ linktoc=all,
    colorlinks, linkcolor={palatinateblue},
    citecolor={brightpink}, urlcolor={amaranth}
}
\begin{document}

\title{Inflationary magnetogenesis from non-minimal coupling in large- and small-field potentials}

\author{Orlando Luongo}
\email{orlando.luongo@unicam.it}
\affiliation{Universit\`a di Camerino, Via Madonna delle carceri 9, 62032 Camerino, Italy.}
\affiliation{Department of Nanoscale Science and Engineering, University at Albany-SUNY, Albany, New York 12222, USA.}
\affiliation{INAF - Osservatorio Astronomico di Brera, Milano, Italy.}
\affiliation{Al-Farabi Kazakh National University, Al-Farabi av. 71, 050040 Almaty, Kazakhstan.}

\author{Antonino Giacomo Marino}
\email{antonino.marino@unicam.it}
\affiliation{Universit\`a di Camerino, Via Madonna delle carceri 9, 62032 Camerino, Italy.}

\author{Tommaso Mengoni}
\email{tommaso.mengoni@unicam.it}
\affiliation{Universit\`a di Camerino, Via Madonna delle carceri 9, 62032 Camerino, Italy.}
\affiliation{INAF - Osservatorio Astronomico di Brera, Milano, Italy.}
\affiliation{Istituto Nazionale di Fisica Nucleare (INFN), Sezione di Perugia, Perugia, 06123, Italy.}

\begin{abstract}
We investigate inflationary magnetogenesis in a scenario where conformal invariance of electromagnetism is broken through a \emph{non-minimal Yukawa-like coupling between the inflaton and the Ricci scalar}. We account for electromagnetic backreaction and the Schwinger effect, analyzing both standard single-field inflation and a generalized K-essence framework, \emph{dubbed quasi-quintessence}. We consider inflationary potentials compatible with Planck satellite constraints, including Starobinsky and $\alpha$-attractor models for large fields, as well as hilltop scenarios for small fields. Moreover, we explore very different functional electromagnetic couplings, introducing a novel ansatz modeled for small-fields. We show that the non-minimal coupling plays a central role in controlling the dynamics, \emph{acting as a timing parameter that regulates the onset of electric backreaction and the Schwinger regime}. This leads to a deep modification of the magnetogenesis process. Indeed, the amplitude of the generated magnetic fields can be enhanced by several orders of magnitude with respect to the minimally coupled case, reaching present-day values up to $B_0 \sim 10^{-13}\,\mathrm{G}$ in large-field scenarios, \emph{which appear as the only ones compatible with observational bounds}. Conversely, small-field models yield negligible magnetic amplitudes and appear non-predictive within our non-minimal framework.
\end{abstract}


\maketitle
\tableofcontents


\section{Introduction} \label{intro}

The origin of the large-scale magnetic fields observed in the Universe remains one of the significant open problems in modern cosmology \cite{giovannini2004}. In particular, a major challenge concerns the generation of magnetic fields possessing very large coherence lengths, exceeding the megaparsec scale. Constraints on the strength of these fields are derived from observations on the cosmic microwave background (CMB) \cite{collaborationPlanck2015Results2016} and from high-energy cosmic rays produced in blazars \cite{tavecchioIntergalacticMagneticField2010,neronovEvidenceStrongExtragalactic2010,andoEvidenceGammaRayHalos2010}. These analyses, together with constraints from the Lyman-$\alpha$ forest \cite{sethiPrimordialMagneticFields2004}, given by the data on absorption lines in the spectra of distant quasars, suggest the magnetic field strength lies within the range $10^{-16} \, \mathrm{G} \lesssim B \lesssim 10^{-10} \, \mathrm{G}$ \cite{martin2008, giovannini2004}. 

Over the years, numerous scenarios have been proposed to explain the origin of cosmic magnetic fields \cite{Widrow_Ryu_Schleicher_Subramanian_Tsagas_Treumann_2012}, broadly divided into astrophysical and primordial mechanisms. Despite significant progress, no consensus has been reached on the dominant paradigm.

For example, astrophysical models, such as the Biermann battery \cite{grasso2001, Attia2021}, generate seed fields that are subsequently amplified by dynamo processes. However, these mechanisms typically lead to coherence lengths of at most kiloparsec scales \cite{davisPrimordialSpectrumGauge2001}, which remain insufficient even in the presence of turbulent amplification \cite{Kulsrud1997b}.

This limitation motivates primordial scenarios, where magnetic fields originate in the early Universe, either during phase transitions or through non-minimal couplings between the inflaton and the electromagnetic sector \cite{durrer2013,giovannini2004, vachaspati1991}.

A widely-studied approach to magnetogenesis relies on breaking the conformal invariance of the electromagnetic sector \cite{vilchinskii2017}, typically through couplings to the inflaton or to curvature invariants. In these scenarios, conformal invariance is restored after inflation, ensuring consistency with standard electromagnetism at late times. Remarkably, within this framework, quantum effects such as Schwinger pair production can significantly affect the evolution of electromagnetic fields \cite{Sobol2018, Hayashinaka2016a, Stahl2016b, Bavarsad2016a, Kobayashi2014}. Depending on the underlying dynamics, the induced current may either screen or enhance the field, effectively modifying the conductivity of the medium. In particular, regimes of negative conductivity have been reported, leading to a transient amplification of the electromagnetic field without requiring strong direct couplings to the inflaton sector \cite{Hayashinaka2016a}.

Motivated by the above considerations, we here investigate magnetogenesis within a conformal symmetry breaking scenario, involving a non-minimally coupled inflationary framework with curvature. In particular, we consider a Yukawa-like interaction between the inflaton and the Ricci scalar \cite{improvement,hertzberg,density,complete,obscons,futamase,futamase1,LUCCHIN1986163}. This non-minimal choice is motivated since such couplings are expected at ultraviolet scales, known to affect several physical processes including, among others, Higgs inflation \cite{higgs, Cheong:2021vdb}, particle production \cite{Ford_2021}, Big Bang Nucleosynthesis \cite{azevedoBigbangNucleosynthesisCosmic2018}, reheating dynamics \cite{preheating} and the spectrum of cosmological perturbations \cite{He:2018mse}. Employing our Yukawa-like term causes the frame issue, \emph{i.e.}, the unclear non-equivalence between the Einstein and Jordan frames \cite{postma,Capozziello_1997,Belfiglio:2024swy}. In this work, we explicitly assess the impact of the aforementioned frame issue on the physical observables of our magnetogenesis scenario, showing that they coincide at leading order in the non-minimal coupling parameter. To do so, the non-minimal coupling constant, $\xi$, is taken small enough to guarantee coincidence between frames. Afterwards, we consider two inflationary setups: standard single-field inflation and a generalized K-essence scenario, dubbed quasi-quintessence (QQ) \cite{D_Agostino_2022,Luongo:2024opv,Carloni:2024ybx,Belfiglio:2023eqi,Belfiglio:2023rxb,Alfano:2025awf}, characterized by a dust-like inflaton with vanishing sound speed \cite{Luongo_2018}. For both cases, we analyze inflationary potentials consistent with Planck observations \cite{Akrami2018}. Specifically, we study both large- and small-field models. In the large-field class, we consider the Starobinsky potential, as well as E- and T-models within the $\alpha$-attractor framework. In the small-field regime, we focus on the quadratic hilltop potential. For the coupling between the inflaton and the electromagnetic sector, we adopt the Ratra model, a power-law extension and a generalized Ratra-like coupling. In the small-field case, we further introduce an \emph{ad hoc} coupling designed to extend magnetogenesis to this regime. Our numerical analysis shows that $\xi$ can significantly enhance the magnetic field amplitude up to $B \sim 10^{-13}\,\mathrm{G}$, corresponding to an increase of about three orders of magnitude with respect to the minimally-coupled case. The coupling strength controls both the energy scale and the evolution of the generated fields, while remaining consistent with current observational constraints. We further impose compatibility with Planck data by requiring agreement with the scalar power spectrum, spectral index, and tensor-to-scalar ratio. We compare our results with previous studies \cite{Sobol_Gorbar_Teslyk_Vilchinskii_2021}, highlighting the role of non-minimal couplings and the differences between large- and small-field scenarios. Finally, we better constrain the coupling parameter within the range $-10^{-3}\lesssim \xi \lesssim 10^{-3}$, \emph{i.e.}, consistent with late-time bounds \cite{Chiba_Kohri_2015}, but in tension with typical values assumed in Higgs inflation \cite{Kaganovich_2026}. We conclude that this suggests viable magnetogenesis scenarios to favor small non-minimal couplings with large-fields only.

The paper is organized as follows. In Sec. \ref{Sec2} we review the main features of the non-minimal coupled inflation, in both the quintessence and QQ frameworks. Sec. \ref{Sez3} is devoted to studying the electrodynamics in the inflationary regime, along with a discussion on the Schwinger effect and the present value of the magnetic field. In Sec. \ref{Sec4}, we introduce different choices for the coupling functions, $f(\phi)$, and for the inflationary potentials, $V(\phi)$, distinguishing between the employment of large or small field models.
Finally, in Sec. \ref{Sec5}, we analyze the numerical outputs deriving from our approaches and, then, Sec. \ref{Sec6} reports conclusions and perspectives.


\section{Non-minimally coupled dynamics}\label{Sec2}

We consider that the inflaton is coupled to a vector field as shown in the action \cite{Sobol2018,giap,Bassett_2006,riotto}
\begin{equation}\label{tot act}
\begin{gathered}
        S=\int\sqrt{-g}\left[\onetwo\MpXi R-\onefour f^2(\phi)\FF+\right.\\+\mathcal{L}_\phi+\mathcal{L}^{ch}(A,\chi)\bigg]d^4x\,,
\end{gathered}
\end{equation}
where $M_P$ is the reduced Planck mass $M_P=1/\sqrt{8\pi G}$ in natural units, $\mathcal{L}_\phi$ is the inflationary Lagrangian, $F_{\mu\nu}=\partial_\mu A_\nu-\partial_\nu A_\mu$ is the \emph{electromagnetic tensor} and $\mathcal{L}^{ch}(A,\chi)$ is the Lagrangian density of the charged and unspecified fields $\chi$ interacting with the vector field through the Schwinger effect. 

The main idea of this formulation is to consider a non-minimal coupling between the electromagnetic sector and the inflaton. Such interaction will cease as inflation ends, restoring the conformal invariance of the electromagnetic theory and, \emph{de facto}, allowing the reheating phase to happen as predicted in the standard picture \cite{Kofman:1997yn}.  

In our scenario, the term $\sim\xi\phi^2R$ plays the role of non-minimal coupling between the inflaton and geometry, as there is no reason \emph{a priori} to neglect such terms throughout the early Universe evolution, considering its high energy scale, see Refs. \cite{Herranen:2014cua, planck}.  

From a geometric viewpoint, the non-minimal coupling leads to modified Einstein field equations, which are written as \cite{Faraoni2000}
\begin{equation}\label{efe}
    \MpXi G_{\mu\nu}=\tilde{T}_{\mu\nu}(\phi,A_\mu)+T^{ch}_{\mu\nu}\,,
\end{equation}
where $G_{\mu\nu}$ is the usual Einstein tensor $G_{\mu\nu}\equiv R_{\mu\nu}-\onetwo Rg_{\mu\nu}$ and the right-hand side features the stress-energy tensors of the fluids populating the Universe. 

Here, $T^{ch}_{\mu\nu}$ refers to those charged particles produced by the Schwinger effect, which are described by $\mathcal{L}^{ch}(A,\chi)$. In particular, since we do not need to directly specify what kind of particles are produced by the Schwinger effect and so we do not need to express $T^{ch}_{\mu\nu}$ explicitly, but rather focus on $\tilde{T}_{\mu\nu}(\phi,A_\mu)$, reading
\begin{align}
    \tilde{T}_{\mu\nu}(\phi,A_\mu)=&T_{\mu\nu}^\phi+\xi\left[g_{\mu\nu}\Box\left(\phi^2\right)-\nabla_\mu\nabla_\nu\phi^2\right]\\
    &-f^2(\phi)g^{\alpha\beta}F_{\mu\alpha}F_{\nu\beta}+\frac{1}{4}g_{\mu\nu}f^2(\phi)F^{\alpha\beta}F_{\alpha\beta}\,\notag,    
\end{align}
where $\Box\equiv\nabla_\alpha\nabla^\alpha$ is the d'Alembertian operator. Specifically, $T_{\mu\nu}^\phi$ is the stress-energy tensor associated with the free inflaton field, while the second term is recovered from the non-minimal coupling to curvature, and the last two from the electromagnetic sector.

The overall dynamics is dictated by the inflationary Lagrangian. In our paper, we focus on two different inflationary scenarios. For our convenience, it is therefore useful to start with the very generic scalar field Lagrangian, written as
    \begin{equation}
\mathcal{L}_\phi = K\left(X,\phi\right) +\lambda\, Y\left[X,\nu\left(\phi\right)\right]-V\left(\phi\right)\,,\label{qq lagr} 
\end{equation}
that includes different types of inflaton fields, once the free functionals are defined.  Precisely, ensuring that  $K(X,\phi)$ is a generalized kinetic term depending on the usual definition $X \equiv\frac12 g^{\alpha\beta}\partial_\alpha \phi \partial_\beta\phi$ and $V(\phi)$ is the potential characterizing the inflationary model, with $\lambda$ a Lagrange multiplier that acts as a constraint on the energy content of the Universe, it is possible to fix the unknown functionals, $K\left(X,\phi\right)$ and $Y\left[X,\nu\left(\phi\right)\right]$, to   explore two main cases: \emph{the standard inflationary scenario} and \emph{a generalization of K-essence inflationary models}. This choice is motivated by the fact that, in the first case, the field usually provides a stiff matter-like behavior, whereas in the second occurrence it reproduces a dust-like contribution \cite{Luongo_2018}. Accordingly, we specify from Eq. \eqref{qq lagr}, the two approaches as follows below. 

\begin{itemize}

\item[1)] {\bf A quintessence inflationary model} defined by
\begin{equation}\label{quint}
    \mathcal{L}_\phi^Q=-\onetwo g^{\mu\nu}\partial_\mu\phi\partial_\nu\phi-V(\phi),
\end{equation}

that can be recovered as a limiting case by setting $K(X,\phi)=-X$ and $\lambda=0$  in Eq. \eqref{qq lagr}. 
Starting from the total action, Eq. \eqref{tot act}, assuming a homogeneous and isotropic Friedmann-Lema\^itre-Robertson-Walker metric, $ds^2=-dt^2+a(t)d\mathbf{x}^2$, and a homogeneous inflaton field, it is possible to obtain inflationary equations of motion. Computing the Klein-Gordon-like equation for $\phi$, we obtain
\begin{equation}
\ddot{\phi}+3H\dot{\phi}+\xi R\phi+V^\prime(\phi)=-\frac{f'(\phi)f(\phi)}{2}\FF,
    \label{KG}
\end{equation}
where the Ricci scalar is given by $R=6(\dot H+2H^2)$, $H=\frac{\dot a}{a}$ is the Hubble parameter, and $V^\prime(\phi)=\frac{\partial V}{\partial\phi}$.

The stress-energy tensor reads
\begin{equation}
    T_{\mu\nu}^\phi=\nabla_\mu\phi\nabla_\nu\phi-\onetwo g_{\mu\nu}\nabla^\sigma\phi\nabla_\sigma\phi-g_{\mu\nu}V(\phi).
\end{equation}
Consequently, we can find the expressions for the inflaton density and pressure, respectively
\begin{align}
        \rho_\phi&=\frac{\dot{\phi}^2}{2}+V(\phi)+6\xi H\phi\dot{\phi}\,,\\
        p_\phi&=\left(\onetwo-2\xi\right)\dot{\phi}^2-V(\phi)-2\xi\phi\ddop-4\xi H\phi\dop\,.
    \end{align}
For this inflationary paradigm, we expect a constant value for the sound speed 
\begin{equation}
    c_s^2\equiv\frac{\partial p_\phi}{\partial X}\left(\frac{\partial \rho_\phi}{\partial X}\right)^{-1}=1\,.
\end{equation}

    \item[2)] {\bf A QQ model}  consists in a generalized K-essence model describing an exotic fluid characterized by a vanishing sound speed, resembling dust \cite{Luongo_2018}.  
Thus, we model inflation through a dust-like field with non-zero pressure or, in other words, a field describing \emph{matter with pressure}. 
Within this scenario, the cosmological constant problem is mitigated by a phase-transition whose metastable stage leads to cosmological inflation, providing a well-defined cancellation mechanism \cite{D_Agostino_2022,Luongo:2024opv}.
This approach naturally combines features of both the \emph{old} \cite{guth} and \emph{chaotic} \cite{LINDE1983177} inflationary paradigms into a unified framework. 

In this framework, we obtain the QQ inflationary dynamics encoded in the continuity equation
\begin{equation}
    \ddot\phi+\frac{3}{2}H\dot\phi+\frac{\xi R\phi}{2}+\frac{\mathcal{V}^\prime(\phi)}{2}=-\frac{f'(\phi)f(\phi)}{4 
    }\FF,\label{infl dyn qq}
\end{equation}
where $\mathcal{V}(\phi) \equiv V(\phi) - K(\phi)$, with $K(\phi)$ that can be set as a constant term \cite{Luongo_2018, D_Agostino_2022, Luongo:2024opv}, namely $K(\phi)=K_0$.

By defining the corresponding effective $4$-velocity as $v_\alpha \equiv \partial_\alpha\phi/\sqrt{2X}$, the energy-momentum tensor for the QQ field becomes
\begin{equation}
\label{eq:no10}
T_{\mu\nu}^\phi = 2X v_\mu v_\nu +\mathcal{V}(\phi) g_{\mu\nu}\,.
\end{equation}
Within this scenario, the energy density and the pressure of the QQ fluid are given by
\begin{subequations}\label{eq:no11}
\begin{align}
\rho_\phi =\,&2X + \mathcal V(\phi)+6\xi H\phi\dot{\phi}\,,\\
\label{eq:no12}
p_\phi =\, &- \mathcal V(\phi)-2\xi\phi\ddop-4\xi H\phi\dop\,.
\end{align}
\end{subequations}
As previously stated, we can explicitly show that the sound speed vanishes for the QQ when $\xi\to0$, implying no perturbations, as matter
\begin{align}
    c_s^2&\equiv\frac{\partial p_\phi}{\partial \rho_\phi}=-\frac{\partial \left(\mathcal V(\phi)\right)}{\partial X}\frac{\partial X}{\partial \left(2X\mathcal{L}_{,X} + \mathcal V(\phi)\right)}=0.
\end{align}

The expected results of the two models are thus different from those of Eq. \eqref{quint} so that the generalized K-essence models may provide different consequences on magnetogenesis and will be explored throughout this manuscript, comparing the two frameworks, in analogy to the study reported in previous literature, see e.g. Refs.  \cite{Carloni:2024ybx, Alfano:2025awf}.

\end{itemize}

From the non-minimal Einstein's field equations, Eq. \eqref{efe}, we can derive the dynamics of the Ricci scalar in the form of modified Friedmann equations, which are written in terms of $R$ and $H$ as
\begin{subequations}
    \begin{align}
    R&=-\frac{\left(1-6\xi\right)\dot{\phi}^2-4V(\phi)-6\xi\phi\ddop-18\xi H\phi\dop}{M_p^2-\xi\phi^2}\,,\label{ricci}\\
    H^2&=\frac{1}{3\MpXi}\left(\rho_\phi+\rho_{EM}+\rho_\chi\right)\,,\label{Fried}
    \end{align}
\end{subequations}
where we defined $\rho_\chi$ as the density of charged particles generated by the Schwinger effect and $\rho_{EM}$ as the generic density contribution of the electromagnetic sector, as we will see later in this work.

\subsection{The role of non-minimal coupling}

At this stage, the dynamics is fully specified once the non-minimal coupling $\xi$ is fixed. The literature covers a wide range, from the conformal value $\xi=1/6$ \cite{futamase1} to large negative couplings $\xi\sim -10^{3}$ \cite{Bezrukov_Shaposhnikov_2009}, down to small values $|\xi|\sim 10^{-3}$ \cite{futamase, preheating}. In a conservative way, it is possible to argue that the weak coupling regime is the best choice for our magnetogenesis picture, 
as larger couplings are expected to induce observable deviations from standard cosmology. This choice is supported by several considerations. In this respect, we enumerate three arguments below. 
\begin{enumerate}
    \item The effective gravitational coupling
\begin{equation}
G_{\rm eff}=\frac{1}{M_P^2-\xi\phi^2},
\end{equation}
is positive definite in order to avoid repulsive effects \cite{preheating}. For field excursions up to $\phi_{\max}\sim 5M_P$, this implies $\xi \lesssim 4\times 10^{-2}$, guaranteeing the background evolution consistency. 

\item Inflationary dynamics further constrains $\xi$. The non-minimal coupling modifies the effective potential and may prevent the inflaton decay \cite{Luongo:2024opv}. For instance, in Starobinsky-like models \cite{starob,1983SvAL....9..302S}, one requires $V'_{\rm eff}(\phi_{\rm in})>0$, implying $\xi \gtrsim -10^{-3}$. Although model-dependent, this bound avoids significant departures from standard inflation.

\item Finally, magnetogenesis imposes a stronger restriction. For $\xi>10^{-3}$, the electric field develops rapid oscillations together with an unphysical growth of its energy density, while a smooth evolution is phenomenologically required. This selects $\xi \lesssim 10^{-3}$.

\end{enumerate}

Combining the above bounds, we can adopt the conservative constraint:
\begin{equation}
|\xi|\lesssim 10^{-3}\,,
\label{boundxi}
\end{equation}
which ensures gravitational stability, preserves the inflationary dynamics and avoids pathological behavior within the electromagnetic sector.


\section{Electrodynamics during inflation} \label{Sez3}

From now on, we express the electromagnetic sector as a function of the electric field by defining\footnote{We define the electric field by the condition $F_{0i}=E_i(t)a(t)$ with $E=|\Vec{E}|$ and, analogously, the magnetic field by $F_{ij}=a^2(t)\epsilon_{ijk}B_k(t)$. All the equations are written with $c=1$.}: $-\onefour f^2(\phi)F^{\mu\nu}F_{\mu\nu}= \rho_E-\rho_B\simeq \rho_E$, effectively assuming that for the evolution of the cosmological background the magnetic field density can be considered subdominant with respect to the electric counterpart. In particular, the assumption $\rho_B\ll\rho_E$ will receive a proper justification by the end of this section. 

As a matter of fact, the enhancement of the electric component turns out to be so effective that it eventually leads $\rho_E$ to reach a comparable magnitude to the inflaton density. The corresponding \textit{backreaction} leads us to include the electric energy density in the equations for the inflationary background.

The evolution of the electric density $\rho_E$ is dictated by Maxwell's equations, which are modified by the presence of the coupling with the inflaton field. According to Eq. \eqref{tot act}, we end up with
\begin{equation}
    \frac{1}{\sqrt{-g}}\partial_\mu\left(\sqrt{-g}f^2(\phi)F^{\mu\nu}\right)=-j^\mu\,,
\end{equation}
where a current source has been introduced, to account for the influence of the charged particle fields determined by 
\begin{equation}
    j^\mu\equiv\frac{\partial\mathcal{L}^{ch}}{\partial A_\mu}\,.
\end{equation}
Following the standard recipe, we  identify $j^\mu$ with the Schwinger current for the production of charged particles in the presence of strong electric fields, where its exact form is explicitly discussed in the next subsection.

The influence of this Schwinger current is, then, consequently included in the continuity equation of the electric field density by the introduction of a source term proportional to the inverse of the scale parameter, as follows
\begin{equation}
    \dot{\rho}_E+4H\rho_E+2\frac{\dot{f}}{f}\rho_E=-\frac{1}{a}\left(\mathbf{E}\cdot\mathbf{j}\right)+\frac{H^3}{4\pi^2}\left[H^2+\left(\frac{\dot{f}}{f}\right)^2\right]\,.
    \label{rhoE}
\end{equation}
Above, we also included the source term given by the electromagnetic modes crossing the horizon as inflation proceeds, represented by the second term on the right-hand side Eq. \eqref{rhoE}. 

Provided that the role of the Schwinger source term is to transfer energy from the electric fluid to the field of charged particles, we require the continuity equation for $\rho_\chi$ density, 
\begin{equation}
    \dot{\rho}_\chi+4H\rho_\chi=\frac{1}{a}\left(\mathbf{E}\cdot\mathbf{j}\right)\,,
    \label{rhoX}
\end{equation}
where $4 H(t) \rho_\chi$ enforces the ultra-relativistic energy scale regime, with the particle masses fulfilling $\frac{m}{H}\ll1$.

In Eq. \eqref{rhoE}, the last term on the right-hand side represents a source contribution describing the modes of the electric field crossing the cosmological horizon. As exhaustively explained in Refs. \cite{Sobol2018, vilchinskii2017}, modes of the electric field are causally connected when the associated wavelength is smaller than the comoving Hubble horizon $d_H=(aH)^{-1}$, or, in other terms, that $k \gg a H$  
\cite{Lyth_Seery_2008}. These sub-Hubble modes have not crossed the horizon yet and exhibit an oscillatory and damped behavior, \emph{de facto} not contributing significantly to the observed value of the electromagnetic field.
On the other hand, the modes that satisfy the condition $k\ll a H$, namely the super-Hubble modes, freeze out after having already escaped the horizon and, at later stages of the Universe's evolution, re-enter the horizon, thereby contributing to the current electromagnetic field as classical long-wavelength contributions.

Numerically, this means that at a given time $t$, the contribution to the electric field density does not consider the modes whose comoving wavenumber is $k>aH$, so the energy density is integrated up to the cut-off defined by the size of the horizon at time $t$
\begin{equation}
\rho_E=\int_{k_i}^{k_H}\frac{d\rho_E}{dk}(t)dk\,,\quad k_H(t)=a(t)H(t)\,.
\end{equation}
Given that the Hubble horizon changes in time, we have that the energy contribution of the crossing modes enters the continuity equation of the electric density by means of the source term
\begin{equation}
\left(\dot\rho_E\right)_H=\left.\frac{d\rho_E}{dk}\right|_{k=k_H}\cdot\frac{dk_H}{dt}\,.
\end{equation}
As suggested in Ref. \cite{martin2008}, a simple formulation for this expression can be obtained by means of the change of variable $\mathcal{A}=afA$, where $A$ is the electromagnetic vector potential. Thus, working in Fourier space with $\mathcal{A}(k,t)$ denoting the mode components, the expression for the electric mode spectrum turns out to be
\begin{equation}\label{pse}
    \frac{\partial\rho_E}{\partial k}=\frac{f^2(\phi)}{2\pi^2}\frac{k^2}{a^2}\left|\frac{\partial}{\partial t}\frac{ \mathcal{A}(k,t)}{f(\phi(t))}\right|^2\,,
\end{equation}
where now the mode components can be computed performing a convenient coordinate transformation from cosmic to conformal time $\eta$, \emph{i.e.}, $d\eta= dt/a(t)$. Hence, the wave equation for the gauge field reads
\begin{equation}\label{ModeEq}
    \frac{\partial^2}{\partial\eta^2}\mathcal{A}(k,\eta)+\left(k^2-\frac{1}{f}\frac{\partial^2f}{\partial\eta^2}\right)\mathcal{A}(k,\eta)=0\,.
\end{equation}
As usually adopted in the literature, during inflation, the above equation can be integrated assuming the existence of a Bunch-Davies vacuum condition\footnote{Alternatives to choosing a Bunch-Davies vacuum can be found in Refs. \cite{Goldstein:2003qf, PhysRevD.110.103502}. } \cite{Bunch:1978yq}, by imposing the asymptotic behavior 
 \begin{equation}\label{BDIC}
     \mathcal{A}(k,t)=\frac{1}{\sqrt{2k}}e^{-ik\eta(t)}\,,\quad k\eta(t)\rightarrow-\infty\,.
 \end{equation}
Equation \eqref{BDIC} is an approximate description of the mode behavior when exiting the cosmological horizon \cite{Sobol2018, vilchinskii2017}. Immediately, this leads to a straightforward evaluation of Eq. \eqref{pse}, which is recovered as the exact Horizon source term defined in the continuity equation for the electric field, Eq. \eqref{rhoE}, namely $\frac{H^3}{4\pi^2}\left[H^2+\left(\frac{\dot{f}}{f}\right)^2\right]$.


\subsection{The Schwinger effect}

The Schwinger effect involves the production of pairs of charged
particles in the presence of strong electric fields. The results obtained for the production rates depend on the
kind of particles that are assumed to be produced in the
process, e.g., scalar, spinor or vector particles, and on the method to compute their rates. Over the years, many complementary methods have been developed to compute the Schwinger current $j$ in Eq. \eqref{rhoE} \cite{Gelis_Tanji_2016, Dunne_Schubert_2005, Smolyansky_Roepke_Schmidt_Blaschke_Toneev_Prozorkevich_1997}. 

Remarkably, all the approaches converge when working in the semiclassical approximation\footnote{The semiclassical limit is defined as the regime where quantum processes can be described by classical trajectories corrected by exponential tunneling factors.}, which corresponds to taking the limits $\frac{eE}{H^2}\gg1$ and $\frac{m^2}{eE}\gg1$, where $m$ is the mass of produced particles and $e$ their charge \cite{Sobol2018}. 

For our purposes, the leading orders govern the production rate \cite{Schwartz}, having
\begin{equation}
    \Gamma(E) \;\propto\; 
\frac{(eE)^2}{4\pi^3} \,
\exp\!\left(-\,\frac{\pi m^2}{eE}\right)\,.\label{PartProdRate0}
\end{equation}
We note that, despite the distinct functional forms of the Schwinger current for bosonic and fermionic production channels, these differences reduce to an overall constant factor in the semiclassical regime \cite{Stahl2016a}. Specifically, the proportionality coefficient in Eq.~\eqref{PartProdRate0} is fixed by the number of degrees of freedom of the underlying scalar or spinor theory.

This justifies adopting a unified semiclassical description, in which Eq.~\eqref{PartProdRate0} provides an effective model for the Schwinger current in our numerical analysis. Accordingly, the source term in Eq.~\eqref{rhoE} can be written as
\begin{equation}
    \frac{1}{a}j_S\cdot E=\frac{g_s}{3\sqrt{2}\pi^3}\frac{e^3}{f^3(\phi)}\frac{\rho_E^{3/2}}{H}
    \exp\left(-\frac{f}{e}\frac{\pi m^2}{\sqrt{2\rho_E}}\right)\,,
    \label{Schw}
\end{equation}
where $g_s=1,2$ accounts for the spin degrees of freedom of bosons and fermions, respectively\footnote{Equation~\eqref{Schw} is derived under the assumption of a constant electric energy density. Even though such a condition cannot be consistently fully maintained in curved spacetime, without introducing unphysical currents and thermodynamical inconsistencies, its functional form remains a widely-accepted approximation even for slowly varying fields \cite{Sobol2018, Kitamoto2018}.}.

In addition, the onset of the Schwinger effect ceases the growth of the electromagnetic sector, \emph{i.e.}, the energy stored in the electric field is efficiently transferred to charged particles, driving $f(\phi)$ toward its asymptotic value. Consequently, \emph{standard electromagnetism is restored, marking the end of magnetogenesis}. Moreover, the electric field is subsequently damped, while magnetic fields evolve under cosmic dilution due to cosmic expansion. Hence, current magnetic amplitudes are computed from their peak values obtained immediately before the Schwinger regime.

As we show below, the Schwinger effect becomes relevant as inflation ends, namely when conformal invariance is thus restored. To this end, we neglect the inflaton fluctuations, which in principle could induce electromagnetic inhomogeneities \cite{Ratra:1991av}. 

Within this simplified setup, we find \emph{a posteriori} that the Schwinger contribution provides only a subleading correction to the overall magnetogenesis process. In this respect, a more complete treatment, incorporating terms beyond the leading one, will be explored in future works.


\subsection{Present value of the magnetic fields}

The current magnetic amplitude can be estimated by integrating the modes that survive cosmic expansion \cite{Sobol2018}. Each mode is characterized by a diffusion timescale,
\begin{equation}
\tau_{\rm diff}=4\sigma L^2\,,
\end{equation}
where $L$ is the physical wavelength and $\sigma$ is the cosmological electrical conductivity \cite{grasso2001}. Accordingly, a given mode survives only if $\tau_{\rm diff}(L_0)>t_0$, with $t_0$ the age of the Universe. 

By imposing this condition and expressing the present length scale as $L_0=L(t_i)a_0/a_i$, one obtains a minimal surviving scale $L_{\rm diff}$. This corresponds to $L_{\rm diff}\sim 1\,{\rm A.U.}$, or equivalently to a comoving cutoff $k_{\rm diff}/a_0\sim10^{-27}\,{\rm GeV}$.

The magnetic energy spectrum is given by
\begin{equation}\label{PSB}
    \frac{d\rho_B}{dk}=\frac{k^4}{2\pi^2 a^4}\left|\mathcal{A}(k,t)\right|^2\,,
\end{equation}
which follows the same structure as Eq.~\eqref{pse}. Adopting the vacuum initial condition of Eq.~\eqref{BDIC}, the present magnetic field amplitude then reads
\begin{align}
     B_0&=\left(\frac{a_e}{a_0}\right)\sqrt{2\int^{k_{\rm diff}}_{a_iH}\frac{dk}{k}\frac{d\rho_B}{d\ln k}}\notag\\
     &=\left(\frac{a_e}{a_0}\right)\sqrt{2\int^{k_{\rm diff}}_{a_iH}\frac{dk}{k}\frac{H^4}{4\pi^2}\left(\frac{k}{a_eH}\right)^2}\notag\\
     &=\frac{1}{2\pi}\frac{k_{\rm diff}}{a_0}\frac{k_*}{a_0}e^N\,,
     \label{B0}
\end{align}
where the spectral scaling is derived assuming that backreaction sets in before the diffusion scale exits the horizon \cite{Sobol2018}. Here, $k_i$ denote the modes that cross the horizon at the onset of inflation, whereas $k_*$ is the pivot scale. The scale factors, $a_e$ and $a_0$, refer to the end of inflation and to the present epoch, respectively.

The number of e-foldings is defined in the usual way,
\begin{equation}
    N=\ln\frac{a_e}{a_i}=\int_{t_i}^{t_e}H\,dt\,,
\end{equation}
with $t_i$ and $t_e$ marking the beginning and end of inflation. To relate inflationary scales to present quantities, we fix the post-inflationary expansion through $N_e=\ln(a_0/a_e)\sim 65$ \cite{German:2022sjd, Luongo:2025ovo}, \emph{i.e.}, a reasonable value that agrees with current expectations \cite{Akrami2018}. 

Equation~\eqref{B0} shows that, once $k_{\rm diff}$ and $k_*$ are fixed, the present magnetic amplitude mostly depends  on the total number of e-foldings. Consequently, models yielding the same $a_e$ predict comparable magnetic strengths.

This result holds only if electric backreaction occurs after the diffusion scale crosses the horizon. If the electric field grows too rapidly, Eq.~\eqref{B0} breaks down. This explains why models with identical e-folding numbers can still produce different asymptotic magnetic amplitudes, as illustrated in Appendix~\ref{appendix plots}.

In this respect, in the numerical results reported in the latter Appendix, we can also find proof of consistency for the assumption of subdominance of the magnetic density with respect to the electric one, e.g. in the plots $\rho_B\ll\rho_E$, with a magnitude gap of roughly three orders, confirming previous results, see e.g. Refs. \cite{demozzi2009, vilchinskii2017, martin2008}.

Nevertheless, the subdominance of the magnetic density in the gravitational background can be evaluated starting from the mode spectra, Eqs \eqref{pse} and \eqref{PSB}. Indeed, the time derivative of the coupling function $f$ notably reads $\dot f \simeq f\dot \phi\ll f$ under the slow roll approximation. Within this assumption and recalling that the significant contribution to the modes comes from super-horizon modes, we find that Eq. \eqref{ModeEq} approximately implies $\ddot {\mathcal{A}}\simeq-H \dot{\mathcal{A}}$. Integrating and considering the ratio between the spectra, we obtain
\begin{equation}
    \frac{d\rho_B}{d \rho_E}=\frac{d\rho_B}{d\ln k}\left(\frac{d\rho_E}{d\ln k}\right)^{-1}
= \frac{k^2}{a^2}\frac{|\mathcal{A}|^2}{f^2}\left|\frac{\partial}{\partial t}\frac{\mathcal{A}}{f}\right|^{-2}\simeq\frac{k^2}{a^2 H^2} \ll 1\,.
\end{equation}

\section{Singling out inflationary scenarios}\label{Sec4}

We here discuss the different choices that can be adopted for the inflationary potential, according to the most popular scenarios considered after Planck measurements \cite{planck}. In the following treatment, we will also describe how to construct the coupling function $f(\phi)$ starting from physical constraints.


\subsection{Choosing the potentials}

For our numerical computation, the inflationary models we are considering can be divided into two main classes.

\begin{itemize}
    \item[-] {\bf \emph{Large field potential.}}
    
Large field inflationary models consider the inflaton field evolution of  Planck mass order. The experimentally most promising model is the Starobinsky potential \cite{starob,1983SvAL....9..302S}.

As an alternative to the Starobinsky model, we underline the $E-$ and $T-$ potentials, within the class of $\alpha$ attractor, firstly obtained in supergravity scenarios \cite{ferrara1,kallosh}.

Below, we summarize the Starobinsky,  $E$ and $T$ models, classifying them by the subscripts $St$, $E$ and $T$, respectively, 
\begin{subequations} \label{larpot}
    \begin{align}
     V_{St}(\phi)=&\Lambda^4\left[1-\exp\left(-\sqrt{\frac{2}{3}}\frac{\phi}{M_P}\right)\right]^2\,,\label{star}\\
     V_E^{(n)}(\phi)=&\Lambda^4\left[1-\exp\left(-\sqrt{\frac{2}{3\alpha_n^E}}\frac{\phi}{M_P}\right)\right]^{2n}\,,\label{E pot}\\
     V_T^{(m)}(\phi)=&\Lambda^4\tanh^{2m}\left(\sqrt{\frac{1}{6\alpha_m^T}}\frac{\phi}{M_P}\right)\,,\label{T pot}
     \end{align}
\end{subequations}
where we adopt $n=2$ to generalize the exponential functional form, while for the sake of simplicity, we set $m=1$.

\item[-] {\bf \emph{Small field potential.}}

Small field class consists of spontaneous symmetry breaking models where the inflaton evolves from an unstable equilibrium point to the minimum of the potential. In this respect, we test two hilltop potentials, corresponding to \cite{Lyth_Seery_2008,linde}
\begin{equation}
    V_{H}^{(p)}(\phi)\simeq\Lambda^4\left[1-\left(\frac{\phi}{\mu}\right)^p\right]\,,\quad \phi<\mu\label{hill pot}
\end{equation}
with $p=2;4$. For values of $\phi\simeq \mu$, $V(\phi)\simeq0$  \cite{Mishra_Sahni_2025}, ensuring the cosmological background to be integrated until the kinetic energy of the inflaton field is completely dissipated in the electric field density\footnote{
In small field models, the inflaton is still expected to reach a potential minimum. Nevertheless, the magnetogenesis evolution is mostly sensitive to the first stages of the inflaton dynamics. For the sake of this investigation, we then assume this specific form of the hilltop minimum to not significantly affect the final value for the generated magnetic fields.}

\end{itemize}

The above inflationary models provide free characteristic parameters, such as the energy scale $\Lambda$ and the set of free constants,  conventionally denoted by $\Theta=\left\{\alpha_E,\,\alpha_T,\,\mu\right\}$. 
These parameters can be constrained using  cosmological parameter bounds, namely from the power spectrum for scalar perturbations, its spectral index and the tensor-to-scalar ratio. Clearly, the introduction of the non-minimal coupling is expected to influence the mathematical expression for these physical observables, albeit as explicitly shown in Appendix \ref{appendix observational}, the first order correction in $\xi$ appears negligible and we can lie on cosmological observations, as stated above.

We thus assume from recent experimental bounds \cite{planck}:

\begin{subequations}
    \begin{align}
        &\Lambda\simeq10^{-3}M_{P},\\
        &\phi_0\simeq5 M_{P},\quad {\rm for\,\, large\,\, fields,}\\
        &\phi_0\ll M_{P},\quad {\rm \,\,for\,\, small\,\, fields.}
    \end{align}
\end{subequations}


\subsection{Choosing the non-minimal functions}

The non-minimal coupling function between the electromagnetic sector and the inflaton is \emph{a priori} unknown. Its form  impacts on  electrodynamics and could lead to unphysical scenarios.

The usual accepted form for $f$ requires a monotonously decreasing trend in the variable $\phi$, tending to an asymptotic constant value of unity at the end of inflation \cite{Sobol2018}. 
This prerogative has the advantage of restoring the conformal invariance at late times, while also avoiding the emerging of the strong coupling problem\footnote{The introduction of the coupling function $f$ provides the effective electric charge, $e_{eff}=ef^{-1}$. Small values of $f$ cause a divergence in $e_{eff}$ leading to the \emph{strong coupling problem} \cite{Sharma_Jagannathan_Seshadri_Subramanian_2017}.}.

In addition, this trend has the benefit of mimicking the monotonous behavior of scale parameter or inflaton density, yielding a physical well-motivated ansatz that can be written differently for large and small fields, as follows.

\begin{itemize}
    \item[-] {\bf \emph{Large field coupling.}} Three different coupling functions are here tested. Two are widely-used in literature, while the last represents a novel approach aiming to generalize the well-known Ratra coupling. The first possible choice for the coupling function is to handle $f\propto a^\alpha$, providing
    \begin{equation}
        f(\phi)=\exp\left[\frac{3\alpha}{4}\left(1+\sqrt{\frac{2}{3}}\frac{\phi}{M_P}-\exp\left[\sqrt{\frac{2}{3}}\frac{\phi}{M_P}\right]\right)\right]\,.\label{Fpot}
    \end{equation}
    Another popular choice is the so-called Ratra function, which historically represents the first proposed form for $f$, representing one of the simplest possible models guaranteeing $f'/f=const$
    \begin{equation}
        f(\phi)=\exp\left[\frac{\beta\phi}{M_P}\right]\,.\label{Fratra}
    \end{equation}
    In Ref. \cite{Sobol2018}, an analysis of these two models has been performed, where the choice of setting $\alpha=15$ and $\beta=-2.5$ appears justified.
    
    Here, we also propose a novel generalized version of the Ratra coupling, given by the function
    \begin{equation}
        f(\phi)=\left[1-(q-1)\frac{\gamma\phi}{M_P}\right]^{\frac{1}{q-1}}\,.\label{Fgratra}
    \end{equation}
    It can be shown that in the limit $q\rightarrow1$ the coupling function takes the form of Ratra coupling with $\beta=-\gamma$. Hence, $q$ is a free parameter that we set to be arbitrarily close to $1$, not to deviate significantly from the exponential trend.

    \item[-] {\bf \emph{Small field coupling.}}    For small inflaton field potentials, defined in Eq. \eqref{hill pot}, the aforementioned conditions on the model imply a different form for the coupling function $f$ than those already introduced. In particular, we will resort to using the following ansatz
    \begin{equation}\label{SymRatra}
        f(\phi)=\exp\left[\beta\sqrt{\phi^2+\delta}\right]+1\,,
    \end{equation}

    where we consider the value of the parameter $\beta$ to be comparable with $\beta$ of Eq. \eqref{Fratra} and $\delta>0$ ensures the continuity of $f'(\phi)$ at $\phi=0$. We justify this specific form for the coupling function directly from the hilltop potential. There, we expect the evolution of the inflaton under the discrete symmetry $\phi\rightarrow-\phi$, having moreover $f\sim1$ at the end of inflation. Similarly to the Ratra coupling, it ensures $f'/f\sim const$, preserving magnetogenesis and  Schwinger effect as we described above.
\end{itemize}


\section{Numerical results}\label{Sec5}

Bearing in mind all the aforementioned ingredients, we here test the viability of our proposed models by means of numerical integrations. The numerical methods will let us scope the evolution of the cosmological background and of the electromagnetic fields, given by Eqs. \eqref{KG}, \eqref{Fried},  \eqref{rhoE}, \eqref{rhoX} and by Eq. \eqref{PSB}. 

Our description is broad since it takes into consideration the above five different inflationary potentials, with four different choices for the coupling function $f$. Every test will be assessed for two different values of $\xi$, corresponding to the limiting cases allowed by Eq. \eqref{boundxi}.

To perform our numerical integrations, we consider the following boundary conditions
\begin{subequations}
    \begin{align}
        &a(0)=1\,,\quad \rho_E(0)=\rho_\chi=\rho_B(0)=\rho_0\,,\\
        &\phi(0)=\phi_0\,,\quad\dop(0)=-\frac{V'(\phi)}{\sqrt{3V(0)}}\,,\ \ \ \ \ \ \ \ 
    \end{align}\label{InitCond}
\end{subequations}
where $\rho_0$ and $\phi_0$ are arbitrary free terms that may be chosen as
    \begin{align}
        &\rho_0=10^{-30}M_P^4,\\
        &\phi_0\sim10^{-1}M_P.
    \end{align}
respectively for large and small fields. In particular, considering a largely subdominant radiation bath at the beginning of inflation, we set the first value above for large fields. Conversely, a similar discussion on the case of small field inflation leads to the $\phi_0$ value above cited for small fields.

\subsection{Large field inflation}

The numerical results for large-field inflation are displayed in Figs. \ref{fig:StarPot}, \ref{fig:EmodPot} and \ref{fig:TmodPot}. Each figure corresponds to one of the inflationary potentials discussed in the text, while the cosmological background and the electromagnetic sector are integrated for the three coupling functions defined in Eqs. \eqref{Fpot}-\eqref{Fgratra}, considering both positive and negative values of the non-minimal parameter $\xi$. The numerical integration is extended up to $t\sim10^7M_P$, which is  long enough to fully explore the inflaton evolution and the growth of the electric and magnetic densities up to their peak values. In a few cases, only the onset of the dilution regime is shown, as in Fig. \ref{fig:EmodPot}, subplots {\it b)} and {\it c)}. Moreover, we enumerate below our main findings. 

\begin{itemize}
\item[-] {\bf Effect of $\xi$ on inflaton evolution.}  
For large-field models, the non-minimal coupling changes the effective slope of the inflaton dynamics and, therefore, shifts the time at which the field decays. In practice, depending on the sign of $\xi$, the inflaton evolution is either anticipated or delayed. This is the most direct effect of having the non-minimal sector, since it determines when the electromagnetic background starts to grow efficiently and when the Schwinger regime becomes relevant. In particular, when $\xi$ is chosen close to the critical condition for which the effective force acting on the inflaton is nearly compensated at the initial stage, the scalar field slows down dramatically. \emph{This prolongs the pre-Schwinger phase and enlarges the interval during which magnetogenesis can operate}.

\item[-] {\bf Electric sector and  backreaction onset.}  
A robust result of our numerics is that the electric density typically reaches a maximum value of order $\rho_{\rm max}\sim10^{-12}M_P^4$, almost independently of the precise functional form of the coupling or of the large-field potential under consideration. This behavior is physically expected, since the electric component cannot grow beyond the energy scale allowed by the inflationary background. Hence, although the early growth of $\rho_E$ is model dependent, its peak value remains bounded by the inflaton energy budget. Moreover, for negative values of $\xi$, the electric sector may become strongly sensitive to the non-minimal Lagrangian term and its peak value can be suppressed by several orders of magnitude. \emph{Interestingly, this suppression does not automatically imply an equally severe suppression of the final magnetic field}.

\item[-] {\bf Magnetic amplification and present-day magnitude.}  As physically expected, the strongest magnetogenesis scenarios are found in the large-field class. First, the present-day magnetic amplitude varies significantly with $\xi$, as shown in Figs. \ref{fig:BXiStar}-\ref{fig:BXiEmod}-\ref{fig:BXiTmod}-\ref{fig:BXiQuad}-\ref{fig:BXiQuar}. Second, among all the tested cases, the most efficient one is the Starobinsky potential combined with the $f\propto a^\alpha$ coupling, for which we obtain a present magnetic strength:  $B_0\sim10^{-13}\,\text{G}$ for $\xi\sim-2.5\cdot10^{-4}$. More generally, $f\propto a^\alpha$ systematically provides the largest amplitudes in the large-field sector, typically ranging between $10^{-17}\,\text{G}$ and $10^{-14}\,\text{G}$. This is crucial, \emph{i.e.}, \emph{these values are the only ones that fall in an observationally interesting range}.

\item[-] {\bf Quintessence versus QQ dynamics.}  
The comparison between quintessence and QQ inflation shows that the QQ framework generally leads to a larger electric background, in some cases by roughly one order of magnitude. \emph{This indicates that the matter-like inflaton dynamics of the QQ scenario can enhance the early electromagnetic amplification}. At the same time, the late-time dilution may become more efficient, depending on the specific potential. Thus, \emph{the difference between quintessence and QQ is quantitatively relevant, but it does not alter the overall conclusion that viable magnetogenesis is achieved only in this class of models}.
 
\end{itemize}

Summing up, large-field inflation provides the only predictive and phenomenologically viable magnetogenesis scenarios, since the non-minimal parameter $\xi$ acts mainly as a timing parameter, regulating the inflaton decay and therefore the onset of the Schwinger effect, while the final magnetic amplitude remains simultaneously controlled by the inflationary potential and the coupling function.

\subsection{Small field inflation}

The numerical results for small-field inflation are displayed in Figs. \ref{fig:HillQuad} and \ref{fig:HillQuar}. In this case, the electromagnetic sector is coupled to the inflaton through the function specifically proposed for the hilltop models, Eq. \eqref{hill pot}. We here work analogously to the large-field case, \emph{i.e.}, the numerical integration follows the full cosmological background together with the electric and magnetic sectors, allowing us to compare the quintessence and QQ frameworks in the presence of non-minimal coupling.

\begin{itemize}
\item[-] {\bf Effect of $\xi$ on inflaton evolution.}  
For small-field inflation, the role of $\xi$ is qualitatively reversed with respect to the large-field case. In particular, a positive non-minimal coupling delays the inflaton evolution instead of accelerating it. The reason is that the hilltop potential has its maximum at the origin, so the non-minimal term acts against the motion of the field toward the minimum. \emph{As a consequence, reheating and the onset of the Schwinger effect are postponed}. 

\item[-] {\bf Electric sector and electromagnetic dilution.} 
Also in the small-field class, the electric and magnetic densities remain bounded by the same inflationary energy scale adopted throughout the analysis and for this reason their absolute magnitudes are not extremely different from those obtained in the large-field case. However, the time evolution is much less efficient for magnetogenesis. In particular, the QQ scenario does not substantially modify the growth stage with respect to standard quintessence, while it tends to produce a stronger dilution of the electromagnetic sector after the end of inflation. Hence, \emph{in small-field models the relevant difference is not the peak growth itself, but the fact that the subsequent evolution washes out the generated field more efficiently}.

\item[-] {\bf Magnetic amplification and present-day magnitude.}  
Remarkably, \emph{our finding shows that small-field inflation always produces much weaker present-day magnetic fields}. \emph{The typical amplitude remains of the order of $B\sim10^{-30}\,\text{G}$, which is many orders of magnitude below the observationally relevant interval}. Hence, even though the model can be integrated consistently and the electromagnetic densities remain under control, the resulting magnetic seed is far too small to represent a viable explanation for cosmic magnetogenesis. This is an indication of the small-field tension in describing early-time epochs. Nevertheless, the hilltop potential vanishes for $|\phi|>\mu$, according to Eq. \eqref{hill pot}. This prescription is required to continue the integration after inflation and to follow the late-time evolution and appears viable. However, even in view of this plausible result, small-field inflation remains inefficient for the production of cosmologically relevant magnetic fields.

\end{itemize}

Summing up, unlike the large-field case, \emph{small-field inflation is not a viable magnetogenesis scenario in the present non-minimal setup}. The effect of $\xi$ is reversed at the level of the inflaton evolution, \emph{the QQ dynamics mainly enhances the dilution stage rather than the amplification stage and the final magnetic field remains negligibly small}. Hence, the comparison between the two classes of models is unambiguous: \emph{large-field potentials can generate observationally interesting magnetic seeds, whereas small-field potentials cannot}.


\section{Final outlooks}\label{Sec6}

In this work, we investigated the mechanism of inflationary magnetogenesis. In particular, we considered the breaking of conformal invariance in the electromagnetic sector, implemented via the coupling function $f(\phi)^2$. More precisely, our analysis was carried out within the framework of non-minimal inflation, where the inflaton field is coupled to the Ricci scalar via a Yukawa-type interaction.

The choice of a non-minimal coupling has been motivated by its wide implications in the early Universe dynamics, including but not limited to Higgs inflation, enhanced particle production, modifications to Big Bang Nucleosynthesis, non-trivial reheating dynamics, and changes in the perturbation spectra. Despite the rich phenomenology, the theoretical status of non-minimal coupling remains debated, particularly concerning the equivalence between the Einstein and Jordan frames. The lack of consensus in the literature on whether these two formulations yield physically indistinguishable predictions led us to critically consider both scenarios when formulating the inflationary observables. 

Hence, our results emphasized that, up to first order in a Taylor expansion around small $\xi$, the physical observables, namely the power spectrum, the spectral index and the tensor-to-scalar ratio,  exhibited frame-independent behavior. In our scheme, this supported the consistency of the approach across formulations and provided confidence in the robustness of the resulting predictions. 

In addition to the frame issue, we constrained the inflaton potential, the Yukawa coupling parameter $\xi$ and ensured compatibility with the latest observational constraints, in agreement with the \textit{Planck} satellite results.

To conduct a thorough investigation, we considered two distinct inflationary paradigms. The first based on the standard scalar field model, characterized by perturbations with a sound speed mimicking stiff matter. The second based on a generalized K-essence scenario, referred to as QQ, which behaves as a dust-like fluid with non-zero pressure and vanishing sound speed, mimicking dust. 

Within both the aforementioned frameworks, we derived the dynamical equations governing the background evolution and the behavior of primordial electromagnetic fields under the influence of non-minimal coupling. These equations were numerically integrated to extract the evolution of the cosmological background, including the electric contribution to address the backreaction problem. 

To this end, we computed the main outcomes derived from the Schwinger effect in damping the electric field via particle-antiparticle pair production. The Schwinger source has explicitly been used within the non-minimal coupling, yielding a magnetic field estimate, valid under additional timing assumptions regarding backreaction. In particular, we presumed that the Schwinger effect provided a leading term, compatible with a uniform and dominant electric field, even in the realm of non-minimal coupling.  

Under these hypotheses, we explored both large and small field inflationary models, selecting those potentials that passed the Planck satellite bounds. 

Particularly, for large field potentials, we tested three different functional forms for the coupling $f(\phi)$: \emph{the original Ratra coupling}, \emph{a generic power-law coupling} and \emph{a generalized extension of the Ratra model}, here proposed as a viable alternative to the previous ones. For each coupling, physical magnetogenesis scenarios were shown, respecting  appropriate ranges of parameters.

Conversely, due to the incompatibility between the conventional coupling functions and small field inflation, we introduced a novel coupling function. Designing this new coupling as a modified version of the Ratra model adapted to the small field context enabled a viable magnetogenesis scenario. In particular, our coupling alleviates the issue that large curvature terms may compete with small fields domains, namely where inflation is thought to occur in the frameworks of small field inflation. Even in this case, we focused on the most viable limits in which the scenarios appear physical. However, we showed that the  hilltop potentials are in strong tension with magnetogenesis, since the small field models here presented provided the weakest strength for the present magnetic fields, missing the observational bounds by several orders of magnitude. This result furnishes a further indication against inflation with small field regimes, in agreement with previous efforts developed in the literature.

Last but not least, we compared our trends,  for each model here investigated, with the results previously analyzed in the literature \cite{Sobol2018}. 
Our numerical results highlighted how the presence of the non-minimal coupling modifies the behavior of the decay of the inflaton field immediately before the onset of the reheating phase. Moreover, we remarked that the QQ frameworks appear significant alternatives to describe inflation adopting a matter-like behavior for the inflaton. Indeed, adopting the QQ model caused the electric background to grow at a faster rate than the quintessence framework and to reach stronger values in the case of Starobinsky and E-model potentials. Remarkably, during the reheating epoch, the QQ model appears to enhance the dilution mechanism of the electromagnetic density in the evolving universe.

Accordingly, our treatment relied on upper bounds on the coupling strength, beyond which the solutions became dynamically unstable. Precisely, choosing a value for $\xi$ in the allowed range has the effect of setting the timescale for the magnetogenesis process, giving us a degree of freedom in tuning when the generated magnetic field starts to undergo cosmic dilution. 

More notably, our results also pointed out that the presence of the non-minimal coupling can lead to an enhancement of the expected present day magnetic strength up to values of $B_0\sim10^{-13}\text{G}$, in accordance with the observational bounds. We stress that this result crucially relies on the timing between the onset of electric backreaction and the horizon crossing of the relevant modes. In particular, the estimate of the present magnetic amplitude holds only if backreaction occurs after the diffusion scale exits the horizon, while an earlier onset would spoil the scaling and significantly reduce the final magnetic strength.

Our findings supported the viability of including non-minimal couplings in the electromagnetic sector, a possibility that remains compatible with the most recent late and early-time datasets.

As a natural extension, we plan to go beyond the massless framework by introducing a massive vector field, leading to a Proca-like electromagnetic sector. This may allow a more direct transfer of inflaton energy into the magnetic component and consequently modify the Schwinger dynamics. We will also investigate possible enhancement mechanisms, including more general couplings in the Lagrangian, such as higher-curvature terms or disformal interactions.

\begin{acknowledgments}
The authors thank Bharat Ratra for insightful discussions on magnetogenesis. OL is grateful to Antonio Capolupo and Aniello Quaranta for useful discussions on related topics. OL and TM acknowledge financial support from the National Institute for Astrophysics (INAF) of Brera and warmly thank Roberto Della Ceca. OL and AGM express their gratitude to Mattia Dubbini for critical discussions on magnetogenesis. 
\end{acknowledgments}

\bibliographystyle{unsrt}
\bibliography{0deposito}

\newpage

\appendix

\section{Early-time observational limits}\label{appendix observational}

We work at the pivot scale $k_*$ where the slow roll approximation holds and consequently we have $k_*= 0.05\,$Mpc$^{-1}$, $\ddot{\phi}\simeq0$ and ${R}^\prime\simeq0$ \cite{Liddle_1994}, along with considering the inflaton potential to dominate. 

We also assume the slow roll regime to kick in well before the effects of backreaction become visible. Phrasing it differently, this  refers to the moment when the effects of the electrical density in the Klein-Gordon-like equation become visible.

This allows us to write the approximated equations for the inflatonary dynamics \cite{Ellis_2015, Riotto:2010jd}
\begin{subequations} \label{slow roll infl}
    \begin{align}
     &3H\dot{\phi}+\xi R\phi+V'\simeq0\,,
    \label{SRKG}\\
     &H^2\simeq\frac{V}{3\left(M_P^2-\xi\phi^2\right)}\,,\\
     &R\simeq\frac{4V}{M^2_P-\xi\phi^2}\simeq12H^2\,.
     \end{align}
\end{subequations}

The above conditions appear quite similar, but generalized for the QQ scenario, yielding the limits
\begin{equation}
\dot\phi^2\ll\mathcal V(\phi)\>\Rightarrow\>\ddot\phi\ll\frac{3}{2}H\dot\phi.\label{2 slowroll}
\end{equation}
In other words, the inflationary magnetogenesis setup is formally the same for both fluids.

Moreover, we handle the best fits of the Planck satellite \cite{planck} and employ the curvature power spectrum for scalar field perturbations $\mathcal{P}_\mathcal{R}$, the spectral index $n_S$ and the tensor-to-scalar ratio $r$. From an experimental viewpoint, all of these quantities are evaluated at the horizon crossing for a fixed pivot scale, $k_*= 0.05\,$Mpc$^{-1}$, in other words, when the associated physical wavelength of fluctuations reaches the Hubble horizon \cite{Cook_2015, Lyth_1999}. Accordingly, this horizon crossing is expected to occur at the beginning of the slow roll phase. 

All these used quantities have been shown in Tab.~\ref{tab:}.

\begin{table}[t]
  \centering
  \begin{tabular}{c|c|c}
    \hline\hline
     & $r$  & $n_s$ \\
    \hline
    Planck TT,TE,EE&&\\+lowEB+lensing & $<0.110$ & $0.9659\pm0.0041$  \\\hline
    Planck TT,TE,EE&&\\+lowE+lensing+BK15 & $<0.061$ & $0.9651\pm0.0041$ \\\hline
    Planck TT,TE,EE&&\\+lowE+lensing+BK15+BAO& $<0.063$& $0.9668\pm0.0037$ \\
    \hline\hline
  \end{tabular}\\
  \begin{tabular}{c|c}
     \hspace{5.25cm}  & \hspace{.3cm} $\ln(10^{10}\mathcal{P}_\mathcal{R})$ \hspace{.3cm} \\
       \hline
        TT,TE,EE+lowE& $3.045\pm0.016$\\
        \hline
        TT,TE,EE+lowE+lensing& $3.044\pm0.014$\\
       \hline\hline
  \end{tabular}
  \caption{Observational bounds from different datasets as reported by Planck mission \cite{planck}.}
  \label{tab:}
\end{table}


\subsection{Non-minimally coupled extensions of the power spectra definition}\label{sect 4.1}

We derive the inflationary observables in the presence of a non-minimal coupling by working in the Jordan frame.

\begin{itemize}

\item[-] {\bf Scalar power spectrum.}  
In the slow-roll regime, the curvature power spectrum reads
\begin{equation}
    \PR=\left(\frac{H^2}{2\pi\dop}\right)^2\,.
\end{equation}
Using Eq. \eqref{slow roll infl}, this can be recast as
\begin{equation}
    \PR=\frac{V}{12\pi^2\MpXi\left[4\xi\phi+\frac{V'}{V}\MpXi\right]^2}\,,
    \label{PR}
\end{equation}
showing explicitly the dependence on the non-minimal coupling.

\item[-] {\bf Spectral index.}  
The scalar spectral index is defined as
\begin{equation}\label{ns}
    n_s=\left.1+\frac{d\ln\PR}{d\ln k}\right|_{k_*}\,.
\end{equation}
Evaluating the derivative at horizon crossing, $k=aH$, and using $d\ln k=Hdt$ together with $d\phi=-H\left[4\xi\phi+\frac{V'}{V}\MpXi\right]dt$, we obtain the corresponding expression for $n_s$.

In the limiting case $\xi\to0$, Eq. \eqref{ns} reduces correctly to
\begin{equation}
n_s|_{\xi=0}=1+2\eta^*-6\epsilon^*\,,
\end{equation}
where the slow-roll parameters are defined as
\begin{align}
    \epsilon&=\frac{M_P^2}{2}\left(\frac{V^\prime(\phi)}{V(\phi)}\right)^2\,,\\
    \eta&=\frac{d\ln\epsilon}{dN}\,.
\end{align}

\item[-] {\bf Tensor sector.}  
Tensor perturbations are described by
\begin{equation}
    \mathcal{P}_T=\frac{8}{ M_P^2}\left(\frac{H}{ 2\pi}\right)^2\,,
\end{equation}
leading to the tensor-to-scalar ratio
\begin{equation}
    r=\frac{\PT}{\PR}=\frac{8\dot\phi^2}{M_P^2H^2}\,.
\end{equation}
In the limiting case $\xi\to0$, one recovers again the standard result $r|_{\xi=0}=16\epsilon^*$.

\item[-] {\bf Weak coupling expansion.}  
Since our analysis is restricted to the regime $|\xi|\ll1$, we expand the observables at first order in $\xi$, obtaining
\begin{widetext}
\begin{subequations}
    \begin{align}
    \mathcal{P}_\mathcal{R}& = \frac{V(\phi)^3}{12 m^6 \pi^2 \left( V'(\phi) \right)^2} - \xi  \frac{V(\phi)^3 \left( 8 \phi V(\phi) - 3 \phi^2 V'(\phi) \right)}{12 m^8 \pi^2 \left( V'(\phi) \right)^3}\,,\label{pr1}\\
    n_s &= 1 + \frac{m^2 \left( -3 \left( V'(\phi) \right)^2 + 2 V(\phi) V''(\phi) \right)}{V(\phi)^2} + \xi  \frac{8 V(\phi)^2 - 10 \phi V(\phi) V'(\phi) + 3 \phi^2 \left( V'(\phi) \right)^2 - 2 \phi^2 V(\phi) V''(\phi)}{V(\phi)^2}\,,\label{ns1}\\
    r &= \frac{8 m^2 \left( V'(\phi) \right)^2}{V(\phi)^2} + \xi  \frac{16 V'(\phi) \left( 4 \phi V(\phi) - \phi^2 V'(\phi) \right)}{V(\phi)^2}\,.\label{r1}
\end{align}
\end{subequations}
\end{widetext}

\end{itemize}

These expressions explicitly show that the leading-order effect of the non-minimal coupling enters as a perturbative correction to the standard inflationary observables.

\section{Plots and numerical computation}\label{appendix plots}

In this appendix, the results for the numerical integration are shown. Precisely, The plots are organized as follows.
\begin{itemize}
    \item[-] Every figure contains the result for a specific inflaton potential. 
    \item[-] The results for each potential are shown twice: both with and without the Schwinger potential term.
    \item[-] Every figure shows three columns, containing the dynamics of the inflaton density $\rho_\phi$, the electric density $\rho_E$ and the magnetic density $\rho_B$. In the case of large field potentials, each row contains instead the results for different coupling functions. With $(a)$ we label the $f\propto a^\alpha$ coupling, with $(b)$ the Ratra coupling and with $(c)$ we indicate the Generalized-Ratra potential.
    \item[-] In each subplot, the gray (black) lines show the result for the QQ (quintessence) field. A straight (dashed) line indicates a negative (positive) value for the $\xi$ parameter.
\end{itemize}

\begin{figure*}
    \centering
    \includegraphics[width=\linewidth]{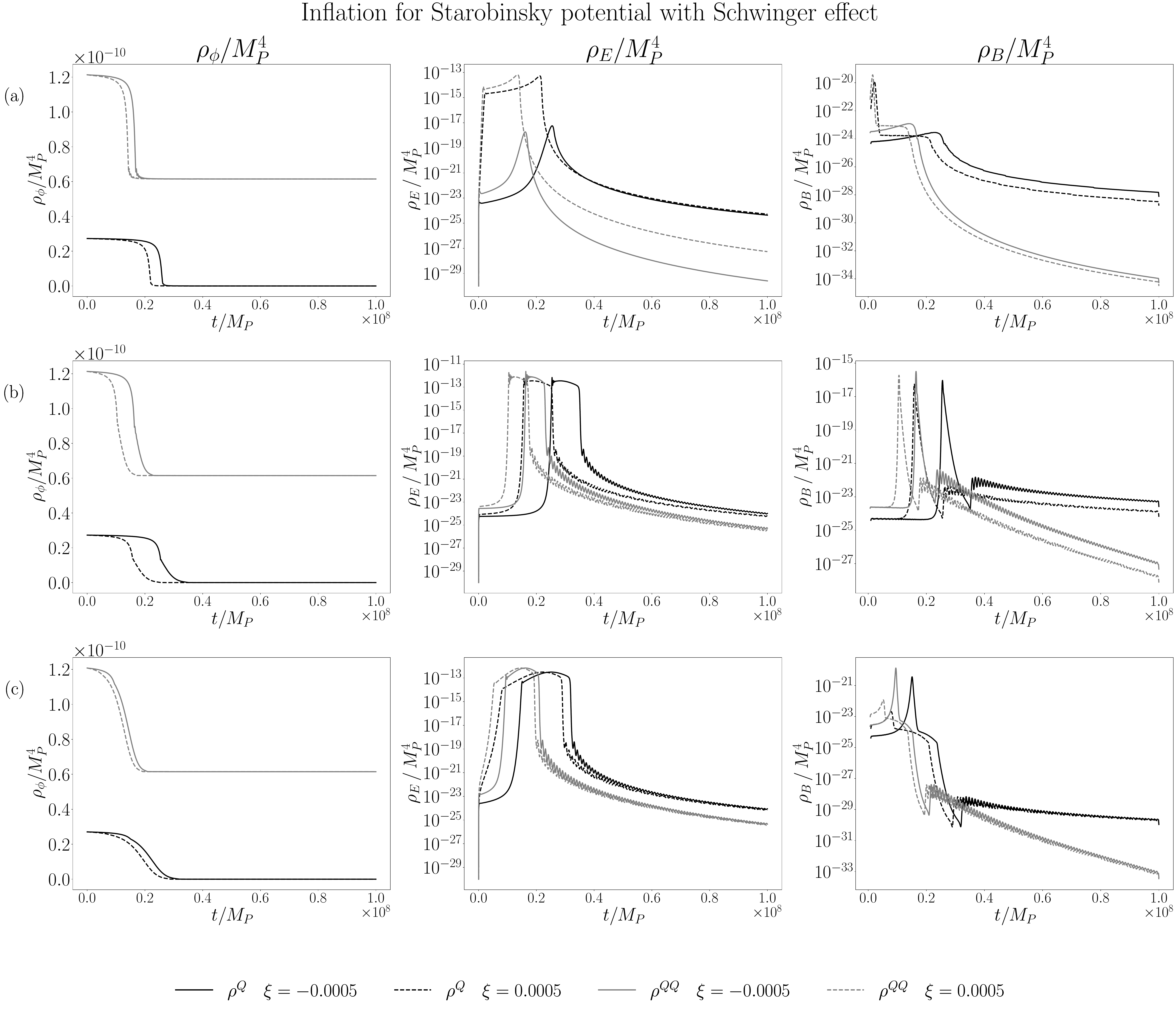}
    \caption{Inflationary densities for Starobinsky potential with Schwinger effect.}
    \label{fig:StarPot}
\end{figure*}

\begin{figure*}
    \centering
    \includegraphics[width=\linewidth]{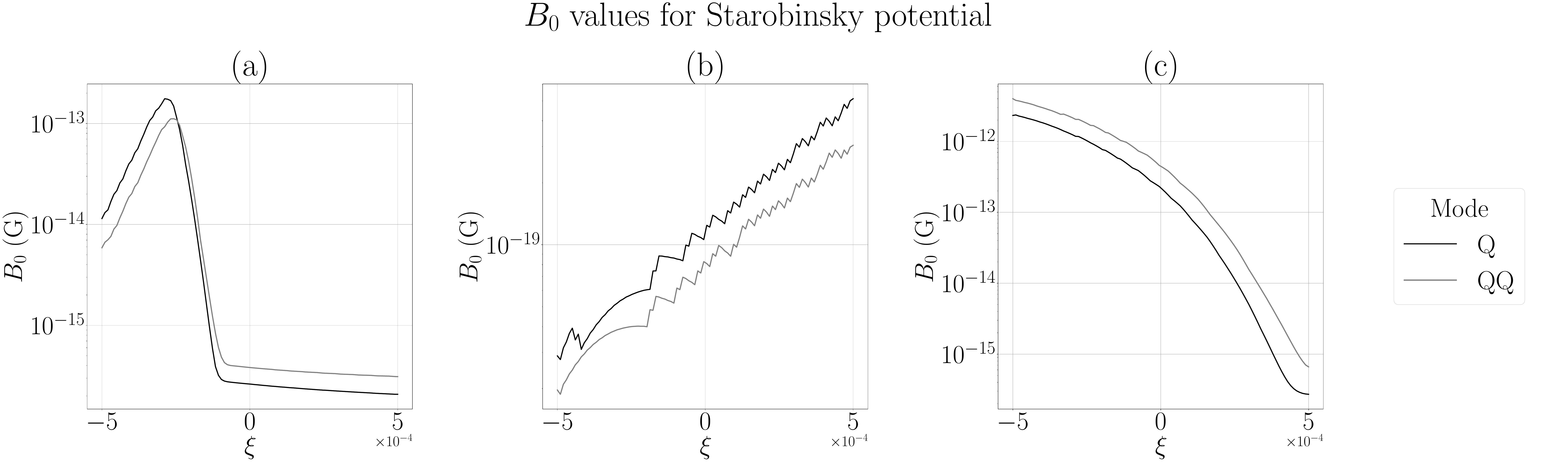}
    \caption{Value of $B_0$ (in Gauss) versus $\xi$ parameter for Starobinsky potential}
    \label{fig:BXiStar}
\end{figure*}

\begin{figure*}
    \centering
    \includegraphics[width=\linewidth]{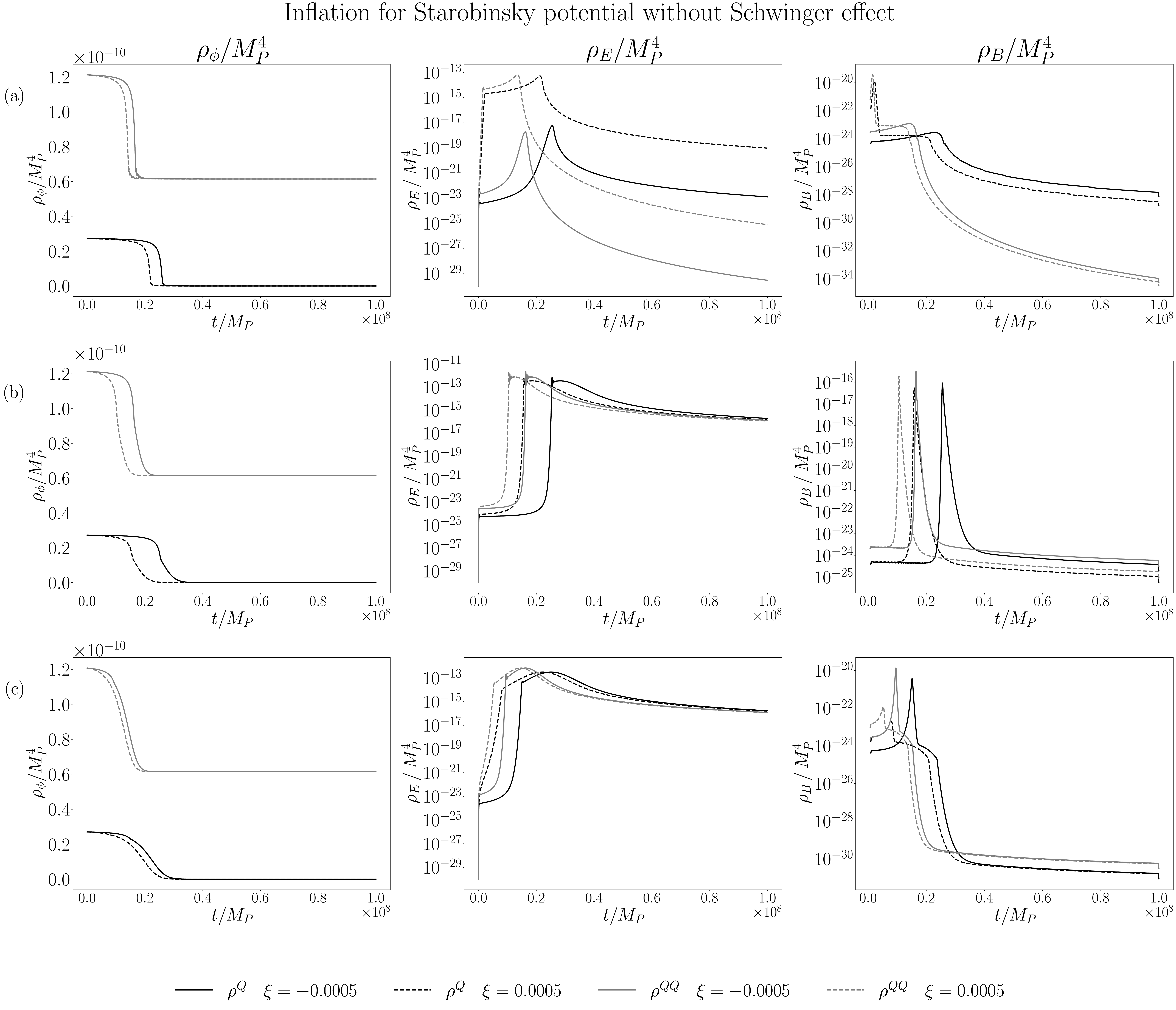}
    \caption{Inflationary densities for Starobinsky potential without Schwinger effect.}
    \label{fig:StarPot1}
\end{figure*}

\begin{figure*}
    \centering
    \includegraphics[width=\linewidth]{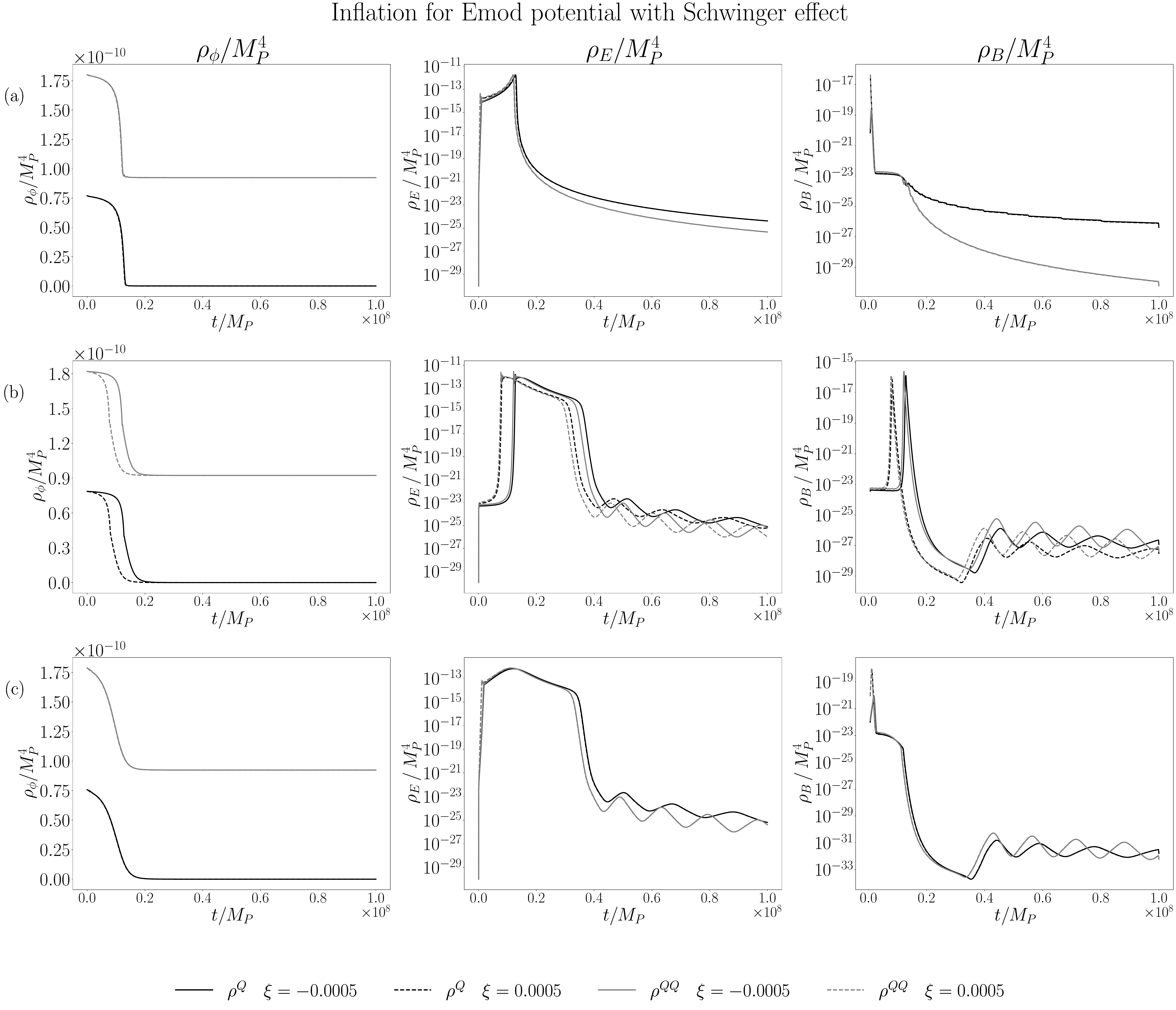}
    \caption{Inflationary densities for Emod potential with Schwinger effect.}
    \label{fig:EmodPot}
\end{figure*}

\begin{figure*}
    \centering
    \includegraphics[width=1\linewidth]{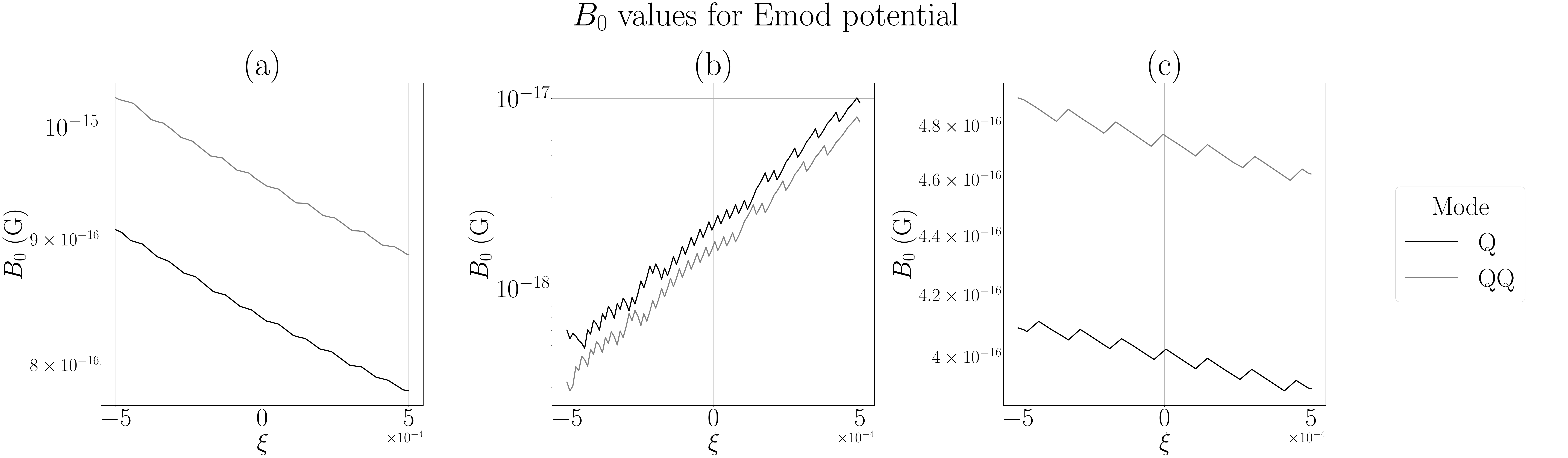}
    \caption{Value of $B_0$ (in Gauss) versus $\xi$ parameter for Emod potential}
    \label{fig:BXiEmod}
\end{figure*}

\begin{figure*}
    \centering
    \includegraphics[width=\linewidth]{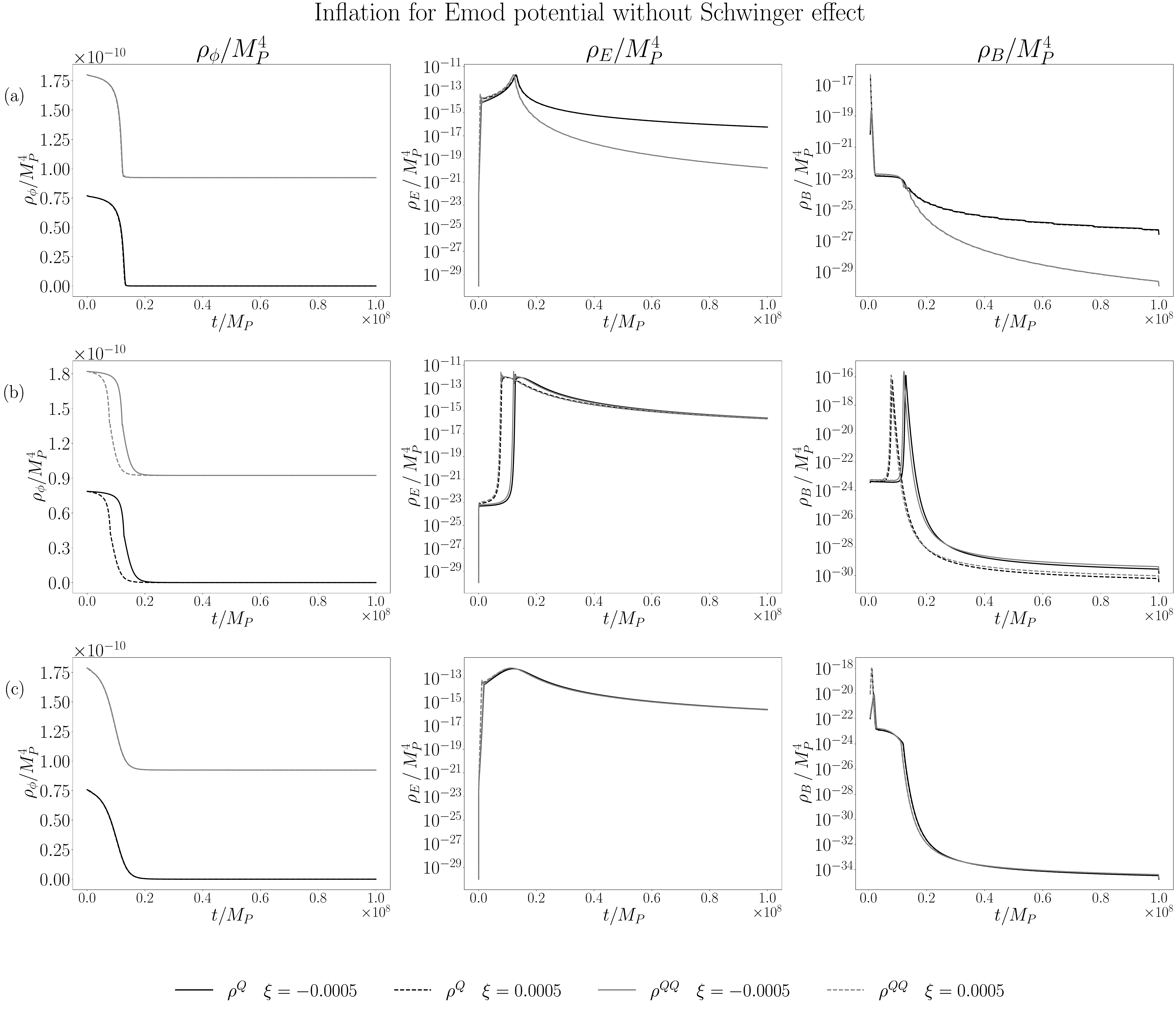}
    \caption{Inflationary densities for Emod potential without Schwinger effect.}
    \label{fig:EmodPot1}
\end{figure*}

\begin{figure*}
    \centering
    \includegraphics[width=\linewidth]{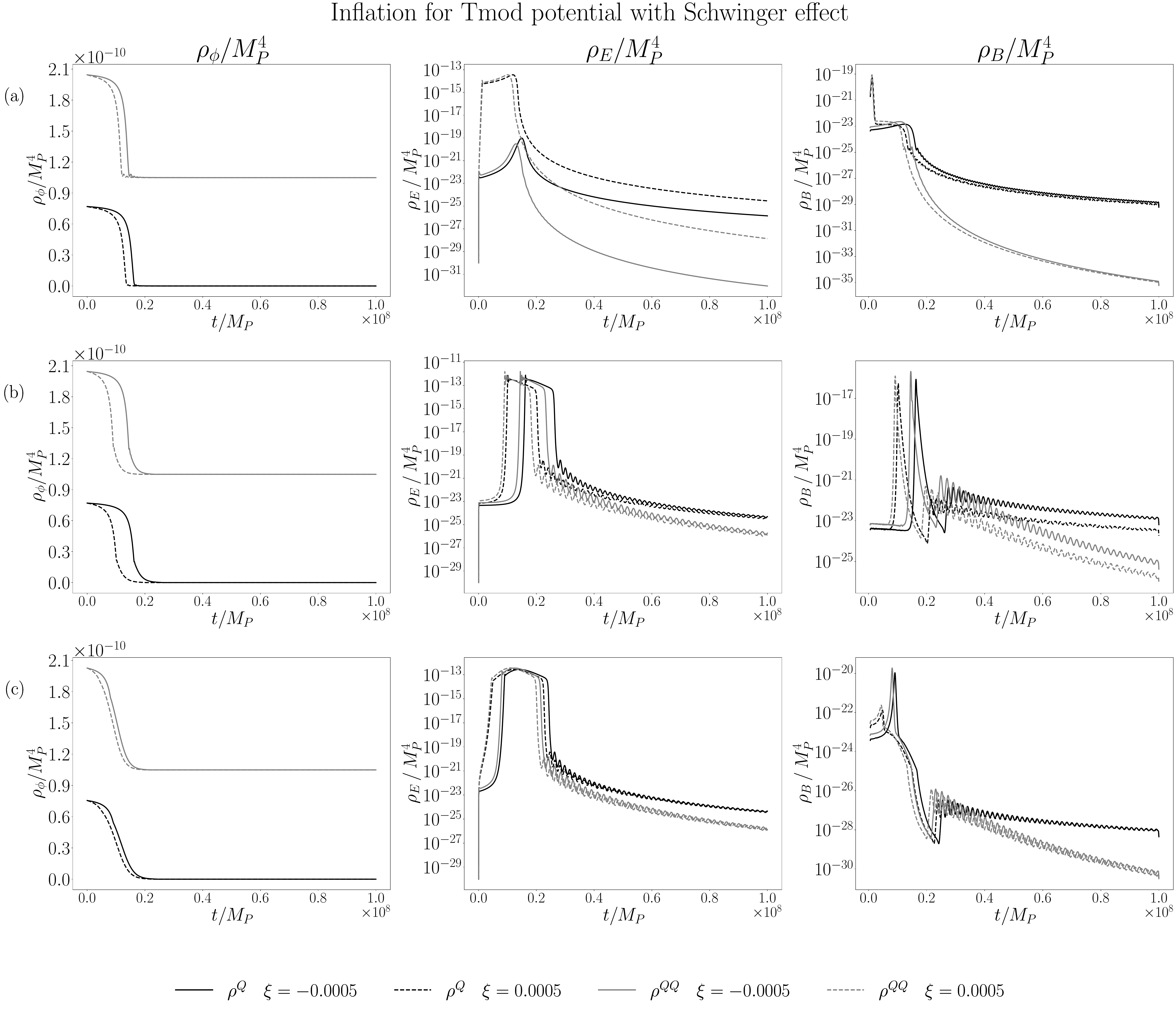}
    \caption{Inflationary densities for Tmod potential with Schwinger effect.}
    \label{fig:TmodPot}
\end{figure*}

\begin{figure*}
    \centering
    \includegraphics[width=1\linewidth]{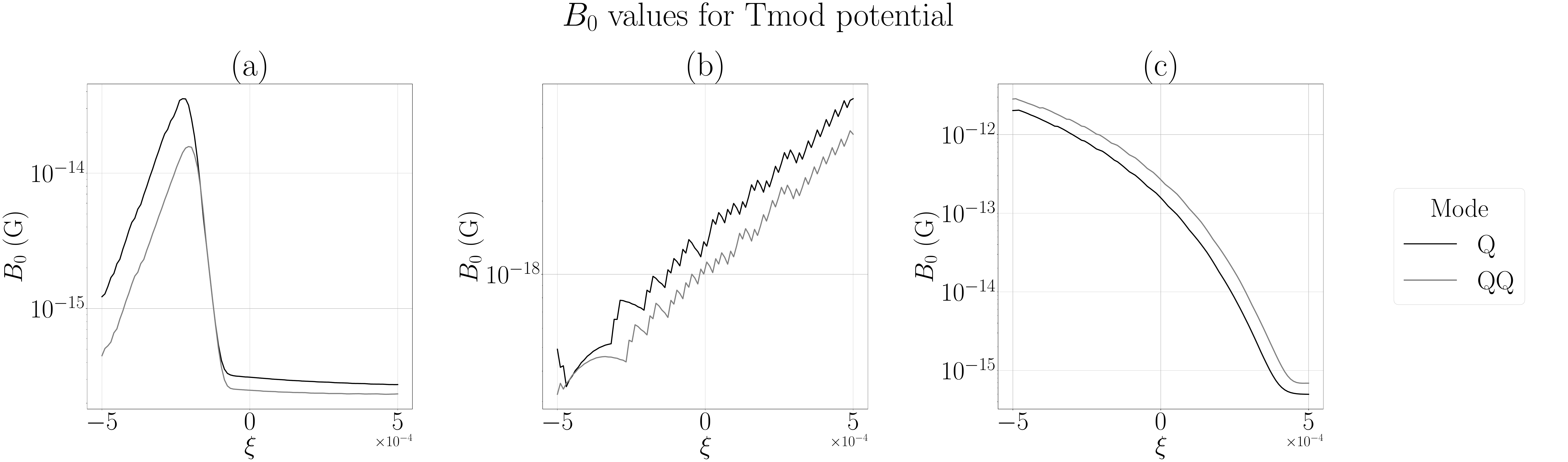}
    \caption{Value of $B_0$ (in Gauss) versus $\xi$ parameter for Tmod potential}
    \label{fig:BXiTmod}
\end{figure*}

\begin{figure*}
    \centering
    \includegraphics[width=\linewidth]{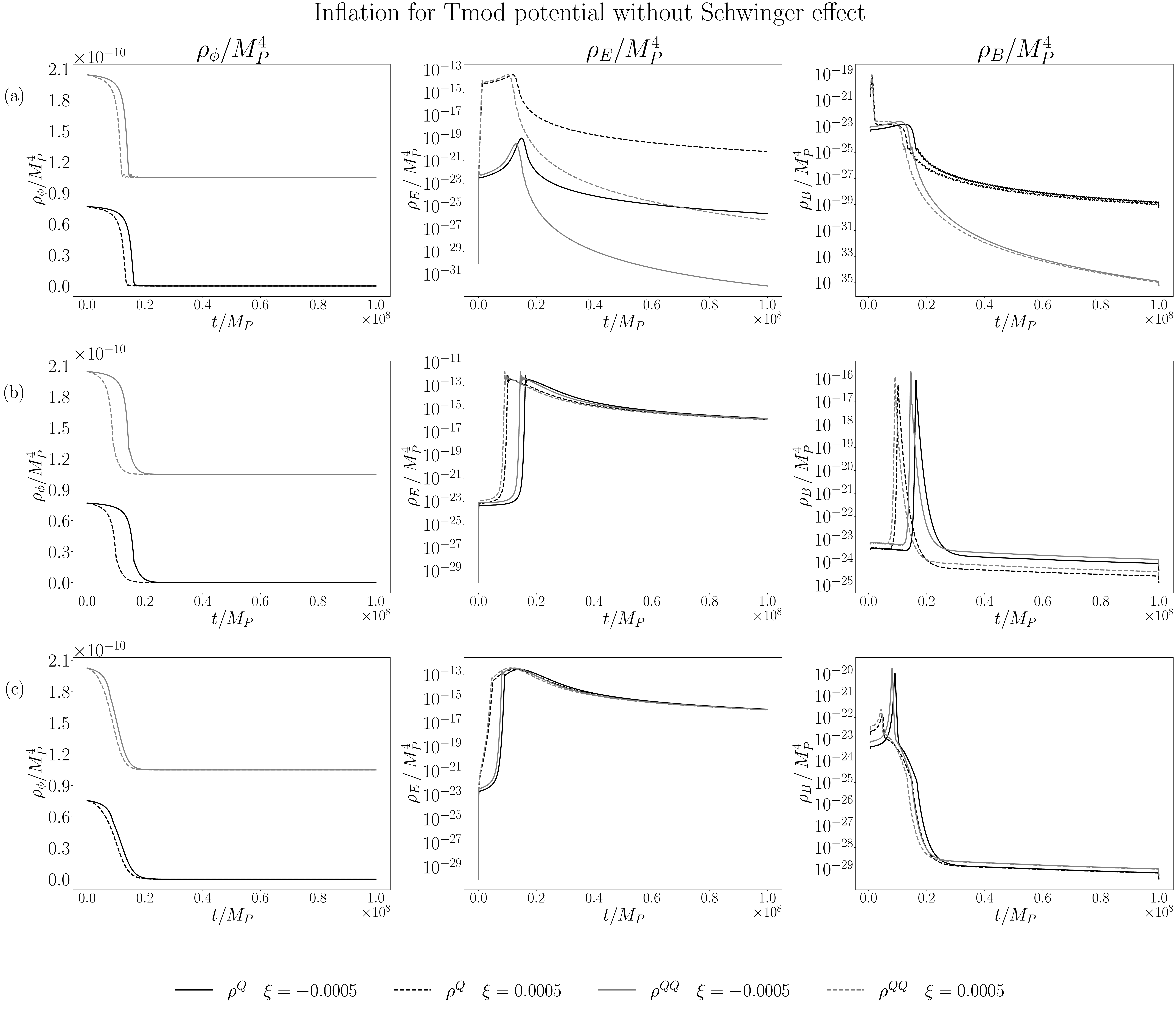}
    \caption{Inflationary densities for Tmod potential without Schwinger effect.}
    \label{fig:TmodPot1}
\end{figure*}

\begin{figure*}
    \centering
    \includegraphics[width=\linewidth]{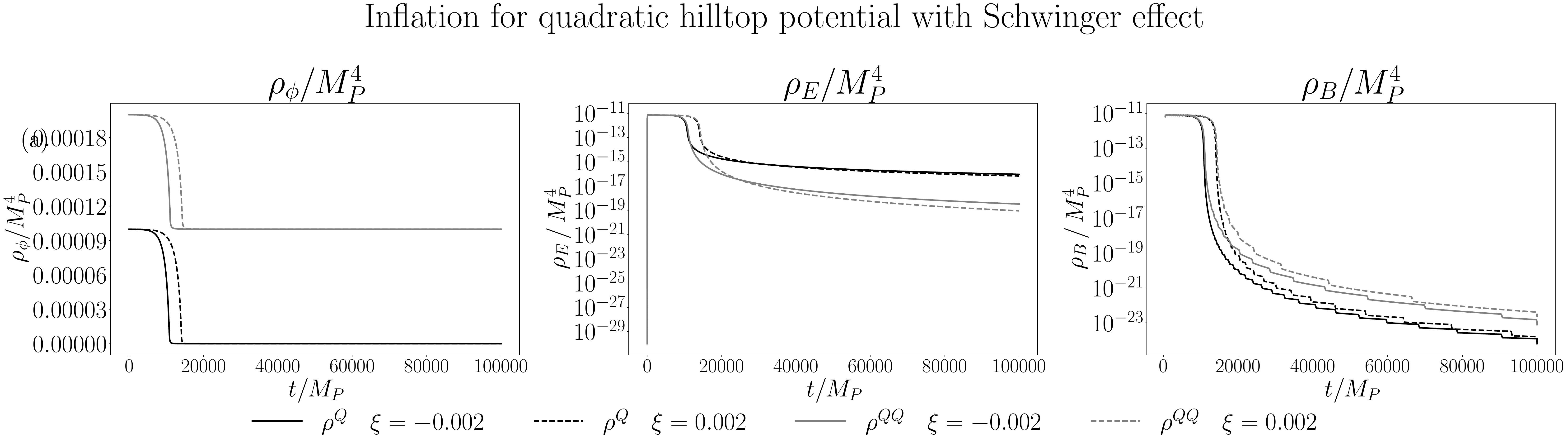}
    \caption{Inflationary densities for hilltop quadratic potential with Schwinger effect.}
    \label{fig:HillQuad}
\end{figure*}

\begin{figure*}
    \centering
    \includegraphics[width=\linewidth]{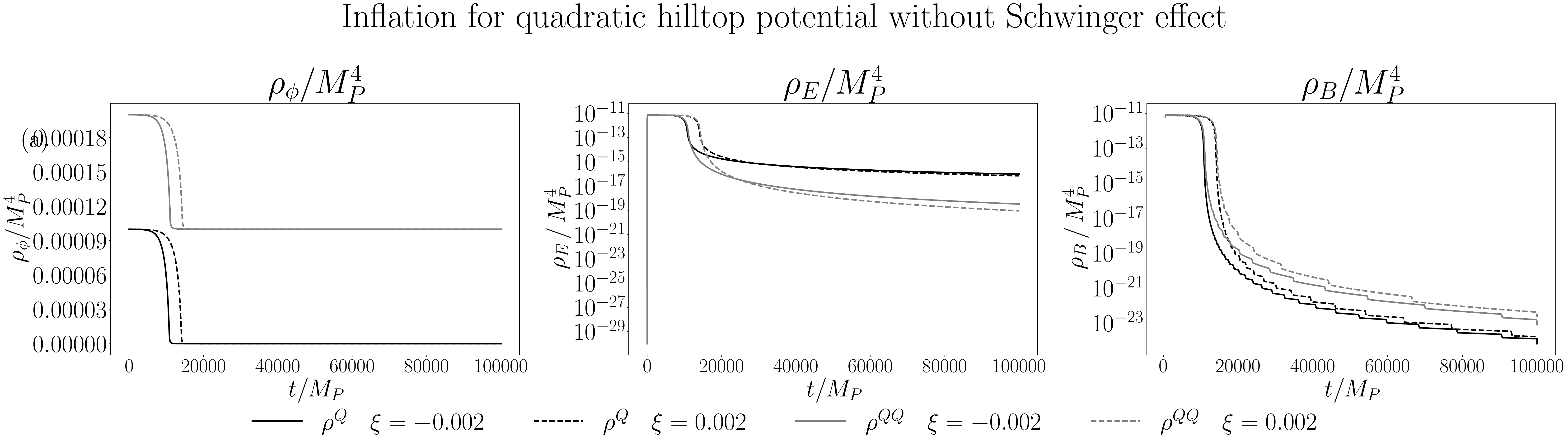}
    \caption{Inflationary densities for hilltop quadratic potential without Schwinger effect.}
    \label{fig:HillQuad1}
\end{figure*}

\begin{figure*}
    \centering
    \includegraphics[width=\linewidth]{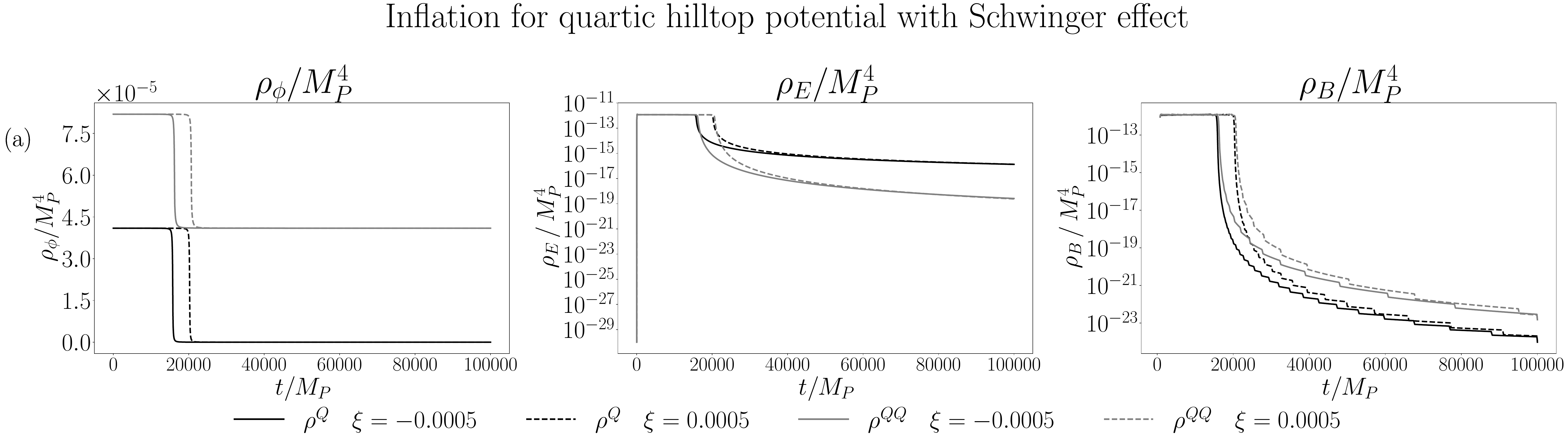}
    \caption{Inflationary densities for hilltop quartic potential with Schwinger effect.}
    \label{fig:HillQuar}
\end{figure*}

Figures \ref{fig:StarPot}-\ref{fig:EmodPot}-\ref{fig:TmodPot}-\ref{fig:HillQuad}-\ref{fig:HillQuar} are the results for each potential considering the presence of the Schwinger effect. 

\begin{figure*}
    \centering
    \includegraphics[width=\linewidth]{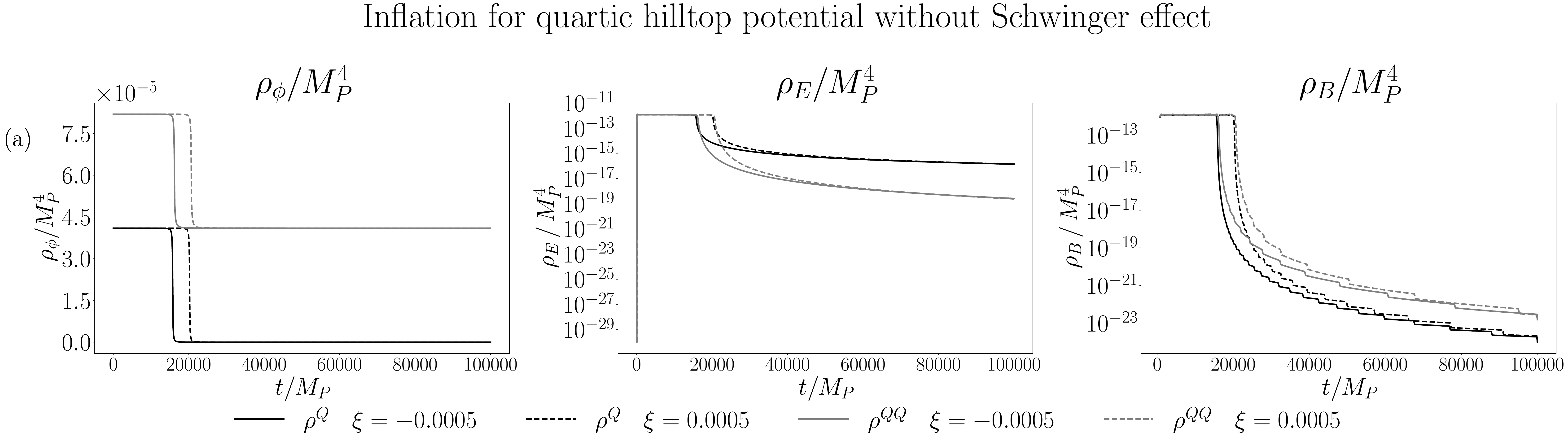}
    \label{fig:HillQuar1}
    \caption{Inflationary densities for hilltop quartic potential without Schwinger effect.}
\end{figure*}

Figures \ref{fig:StarPot1}-\ref{fig:EmodPot1}
\ref{fig:TmodPot1}-\ref{fig:HillQuad1}-\ref{fig:HillQuar1}
 do not consider the presence of the Schwinger effect. 

\begin{figure}
    \centering
    \includegraphics[width=1\linewidth]{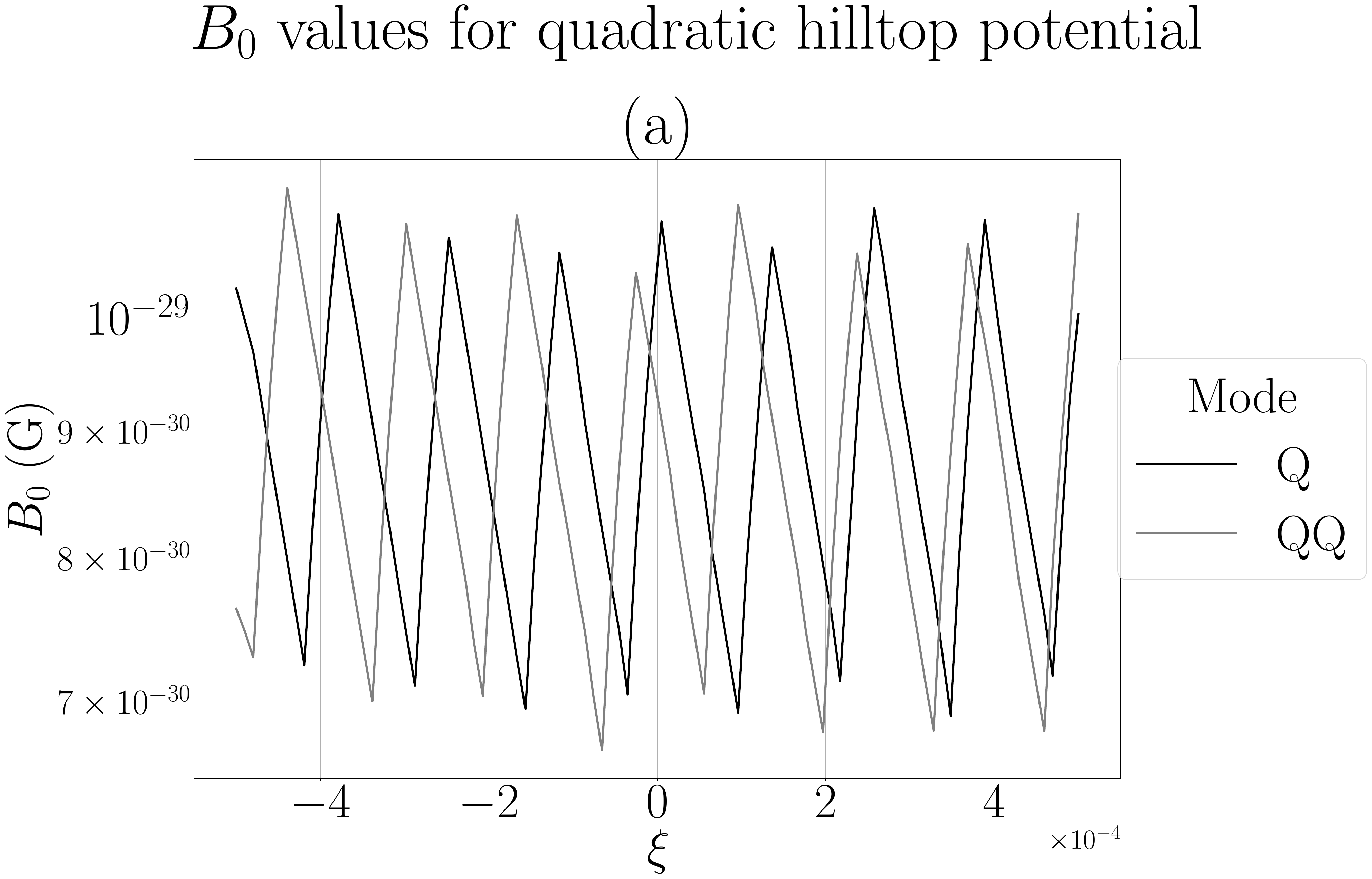}
    \caption{Value of $B_0$ (in Gauss) versus $\xi$ parameter for hilltop quadratic potential.}
    \label{fig:BXiQuad}
\end{figure}

\begin{figure}
    \centering
    \includegraphics[width=1\linewidth]{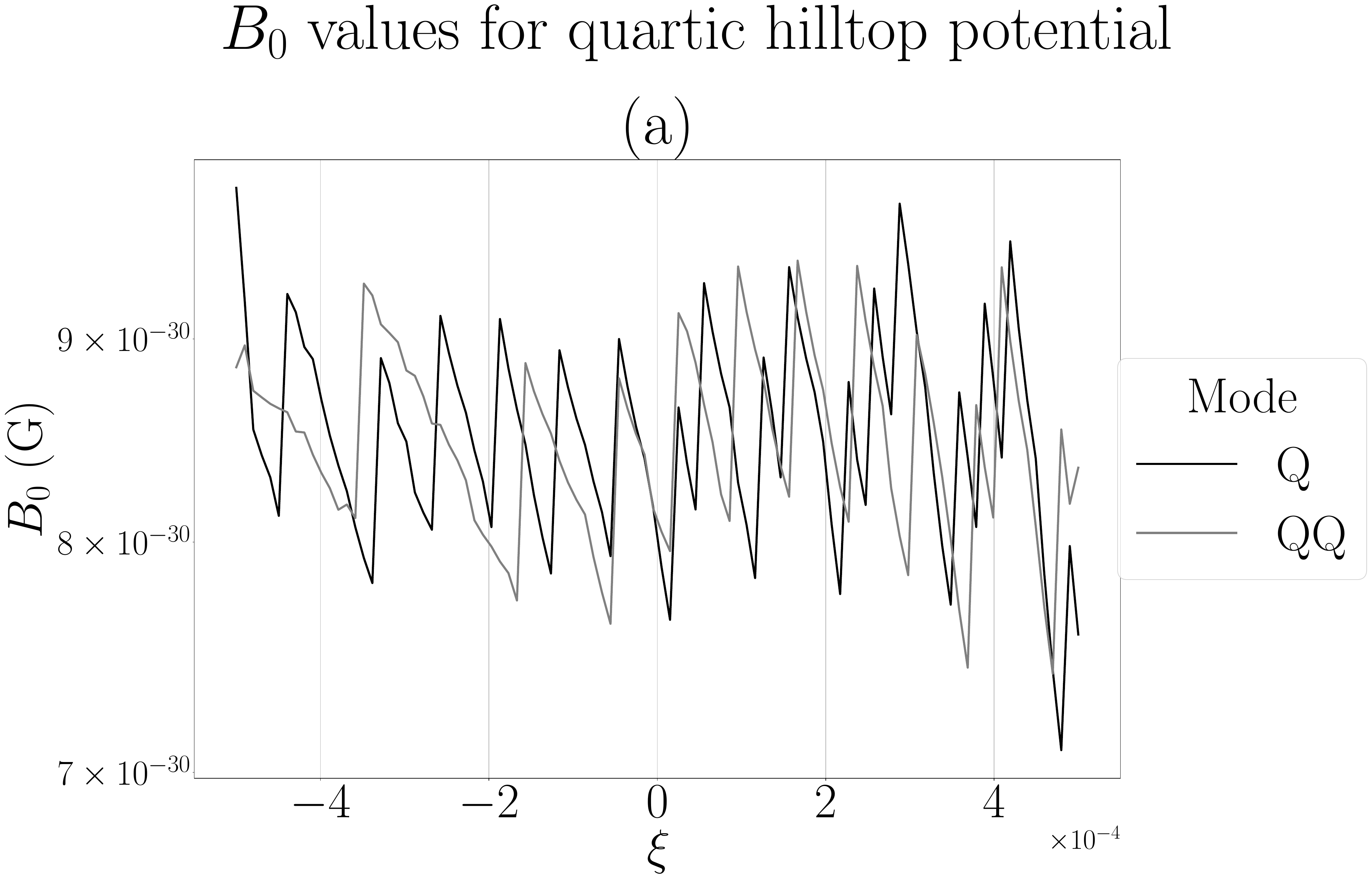}
    \caption{Value of $B_0$ (in Gauss) versus $\xi$ parameter for hilltop quartic potential.}
    \label{fig:BXiQuar}
\end{figure}

In Figs. \ref{fig:BXiStar}-\ref{fig:BXiEmod}-\ref{fig:BXiTmod}-\ref{fig:BXiQuad}-\ref{fig:BXiQuar}, we show the results for the expected nowadays value of the magnetic field (expressed in Gauss units) versus the value of the $\xi$ parameter. 

Importantly, we highlight that the periodic trend that appears in some of these figures is supposedly a numerical artifact due to the momentum space discretization.





\end{document}